\documentstyle[12pt]{article}
\def\journal{\topmargin .175in	\oddsidemargin 0in
	\headheight 0pt	\headsep 0pt
	\textwidth 6.625in % 1.2 preprint size  %6.5in
	\textheight 8.5in % 1.2 preprint size 9in
	\marginparwidth 1.5in
	\parindent 2em
	\parskip .5ex plus .1ex		\jot = 1.5ex}
\journal

\catcode`\@=11
\def\marginnote#1{}
\newtoks\@stequation

\def\subequations{\refstepcounter{equation}%
  \edef\@savedequation{\the\c@equation}%
  \@stequation=\expandafter{\theequation}%   %only want \theequation
  \edef\@savedtheequation{\the\@stequation}% % expanded once
  \edef\oldtheequation{\theequation}%
  \setcounter{equation}{0}%
  \def\theequation{\oldtheequation\alph{equation}}}

\def\endsubequations{\setcounter{equation}{\@savedequation}%
  \@stequation=\expandafter{\@savedtheequation}%
  \edef\theequation{\the\@stequation}\global\@ignoretrue
  \vspace*{-12pt} \\
  }

\@addtoreset{equation}{section}

\def\theequation{\thesection.\arabic{equation}}

\catcode`\@=11
\def\section{\@startsection {section}{1}{0pt}{-3.5ex plus -1ex minus
 -.2ex}{2.3ex plus .2ex}{\raggedright\large\bf}}
\def\l@part#1#2{\addpenalty{\@secpenalty}
 \addvspace{.5em plus 1pt} \begingroup
 \@tempdima 3em \parindent \z@ \rightskip \@pnumwidth \parfillskip
-\@pnumwidth
 {\large \bf \leavevmode #1
\hfil \hbox to\@pnumwidth{\normalsize\rm\hss #2}}\par
 \nobreak \endgroup}
\def\l@section#1#2{\addpenalty{\@secpenalty}
\@tempdima 1.5em \begingroup
 \parindent \z@ \rightskip \@pnumwidth
 \parfillskip -\@pnumwidth
 \bf \leavevmode \advance\leftskip\@tempdima \hskip -\leftskip #1\nobreak\hfil
\nobreak\hbox to\@pnumwidth{\rm\hss #2}\par
 \endgroup}
\newskip\humongous \humongous=0pt plus 1000pt minus 1000pt

\newif\ifdtup

\def\R{{\rm I\!R}}
\def\one{{\mathchoice {\rm 1\mskip-4mu l} {\rm 1\mskip-4mu}
{\rm 1\mskip-4.5mu l} {\rm 1\mskip-5mu l}}}
\def\Q{{\mathchoice
{\setbox0=\hbox{$\displaystyle\rm Q$}\hbox{\raise 0.15\ht0\hbox to0pt
{\kern0.4\wd0\vrule height0.8\ht0\hss}\box0}}
{\setbox0=\hbox{$\textstyle\rm Q$}\hbox{\raise 0.15\ht0\hbox to0pt
{\kern0.4\wd0\vrule height0.8\ht0\hss}\box0}}
{\setbox0=\hbox{$\scriptstyle\rm Q$}\hbox{\raise 0.15\ht0\hbox to0pt
{\kern0.4\wd0\vrule height0.7\ht0\hss}\box0}}
{\setbox0=\hbox{$\scriptscriptstyle\rm Q$}\hbox{\raise 0.15\ht0\hbox to0pt
{\kern0.4\wd0\vrule height0.7\ht0\hss}\box0}}}}
\def\C{{\mathchoice
{\setbox0=\hbox{$\displaystyle\rm C$}\hbox{\hbox to0pt
{\kern0.4\wd0\vrule height0.9\ht0\hss}\box0}}
{\setbox0=\hbox{$\textstyle\rm C$}\hbox{\hbox to0pt
{\kern0.4\wd0\vrule height0.9\ht0\hss}\box0}}
{\setbox0=\hbox{$\scriptstyle\rm C$}\hbox{\hbox to0pt
{\kern0.4\wd0\vrule height0.9\ht0\hss}\box0}}
{\setbox0=\hbox{$\scriptscriptstyle\rm C$}\hbox{\hbox to0pt
{\kern0.4\wd0\vrule height0.9\ht0\hss}\box0}}}}

\font\fivesans=cmss10 at 4.61pt
\font\sevensans=cmss10 at 6.81pt
\font\tensans=cmss10
\newfam\sansfam
\textfont\sansfam=\tensans\scriptfont\sansfam=\sevensans\scriptscriptfont
\sansfam=\fivesans
\def\sans{\fam\sansfam\tensans}
\def\Z{{\mathchoice
{\hbox{$\sans\textstyle Z\kern-0.4em Z$}}
{\hbox{$\sans\textstyle Z\kern-0.4em Z$}}
{\hbox{$\sans\scriptstyle Z\kern-0.3em Z$}}
{\hbox{$\sans\scriptscriptstyle Z\kern-0.2em Z$}}}}

\mathchardef\endbar="375

\def\ceilfill{$\raise3pt\hbox{$\mathsurround=0pt\mathord\endbar$}
  \mkern-2mu \xleaders\hbox{$\mkern-5mu
  \mathord-\mkern-5mu$}\hfill\mkern-7mu
  \raise3pt\hbox{$\mathsurround=0pt\mathord\endbar$}$}

\def\floorfill{$\raise9pt\hbox{$\mathsurround=0pt\mathord\endbar$}
  \mkern-2mu \xleaders\hbox{$\mkern-5mu
  \mathord-\mkern-5mu$}\hfill\mkern-7mu
  \raise9pt\hbox{$\mathsurround=0pt\mathord\endbar$}$}

\def\overcontract#1{\mathop{\vbox{\ialign{##\crcr\noalign{\kern3pt}
  \ceilfill\hskip6pt\crcr\noalign{\kern3pt\nointerlineskip}
  $\hfil\displaystyle{#1}\hfil$\crcr}}}}

\def\undercontract#1{\mathop{\vtop{\ialign{##\crcr
  $\hfil\displaystyle{#1}\hfil$\crcr\noalign{\kern3pt\nointerlineskip}
  \floorfill\hskip6pt\crcr\noalign{\kern3pt}}}}}
\def\b{\beta}
\def\g{\gamma}
\def\d{\delta}

\def\vare{\varepsilon}
\def\m{\mu}
\def\n{\nu}
\def\t{\tau}
\def\p{\pi}
\def\ps{\psi}
\def\r{\rho}
\def\s{\sigma}
\def\l{\lambda}

\def\et{\eta}
\def\L{\Lambda}

\def\Ga{\Gamma}

\def\O{\Omega}

\def\be{\bar{e}}

\def\bE{\bar{E}}

\def\ts{\tilde{\s}}

\def\sL{{\cal L}}
\def\J{{\cal J}}
\def\F{{\cal F}}
\def\E{{\cal E}}

\def\H{{\cal H}}
\def\W{{\cal W}}

\def\G{{\cal G}}
\def\cD{{\cal D}}

\def\C{{\cal C}}
\def\cO{{\cal O}}

\def\N{{\cal N}}

\def\het{\hat{\et}}

\def\hg{\hat{g}}
\def\hga {\hat{\g}}
\def\hH {\hat{H}}

\def\hO {\hat{\O}}

\def\hT {\hat{T}}
\def\hHa {\hH^{adj}}
\def\Ha {H^{adj}}
\def\hB{\hat{B}}
\def\bul{$\bullet\;\,$}
\def\ys{{y_*}}

\def\pa{\partial}
\def\bpa{\bar{\partial}}
\def\ps{\pa_\s}
\def\pt{\pa_\t}

\def\ra{\rightarrow}
\def\lra{\leftrightarrow}

\def\xx{\hbox{ }^*_*}

\def\Tr{{\rm Tr}}
\def\hTr{\hat{\Tr}}

\catcode`\@=12
\begin{document}
\begin{titlepage}

%Draft \# 56  \hfill  \today
\noindent
June 1995     \hfill        UCB-PTH-95/23 \\
hep-th/9506199 \hfill     LBL-37455 \\
% \hfill

\begin{center}
\vskip .05in

{\large \bf The WZW Model \\
 as a Dynamical System on Affine Lie Groups}
\footnote{This work was
supported in part by the Director, Office of
Energy Research, Office of High Energy and Nuclear Physics, Division of
High Energy Physics of the U.S. Department of Energy under Contract
DE-AC03-76SF00098 and in part by the National Science Foundation under
grant PHY90-21139.}

\vskip .2in
K. Clubok\footnote{e-mail: CLUBOK@PHYSICS.BERKELEY.EDU}
and M.B. Halpern\footnote{e-mail: MBHALPERN@LBL.GOV}
\vskip .05in
{\em  Department of Physics, University of California\\
      and\\
      Theoretical Physics Group, Lawrence Berkeley Laboratory\\
      Berkeley, California 94720 \\
       USA}
\vskip .03in

\end{center}
\vskip .1in
\begin{abstract}
Working directly on affine Lie groups, we construct several new
formulations of the WZW model.
In one formulation WZW is expressed as a one-dimensional mechanical system
whose variables are coordinates
on the affine Lie group.
When written in terms of the affine group element, this
formulation exhibits a two-dimensional WZW term.
In another formulation WZW is written as a two-dimensional field
theory, with a three-dimensional WZW term, whose fields are coordinates
on the affine group.
On the basis of these equivalent
formulations, we develop a translation dictionary
in which the new formulations on the affine Lie group are understood as
mode formulations of the conventional WZW formulation on the Lie group.
Using this dictionary,
we also
express WZW as a three-dimensional field theory on the Lie
group with a four-dimensional WZW term.
\end{abstract}
\end{titlepage}
\renewcommand{\thepage}{\roman{page}}
\setcounter{page}{1}
\tableofcontents
\newpage

\renewcommand{\thepage}{\arabic{page}}
\setcounter{page}{1}
\setcounter{footnote}{1}

\section{Introduction}

Affine Lie algebra, or current algebra on $S^1$, was discovered
independently in mathematics \cite{KM} and physics \cite{BH}.
The affine-Sugawara
constructions [2-4]
were the first and simplest conformal field theories
constructed from the currents of the affine algebras. The
WZW model \cite{Nov,Wit}, formulated on Lie groups, is the
world-sheet
description of the general affine-Sugawara construction.
 See Ref.~\cite{Review} for a more detailed history of
affine Lie algebra and conformal field theory.

Affine Lie groups, which are generated by the affine algebras,
are infinite dimensional generalizations of Lie groups.
This paper discusses equivalent reformulations
of the WZW model in terms of
coordinates on the corresponding affine Lie groups.
These reformulations of WZW theory are
based on recent work by Halpern and Sochen \cite{tor}, who used the
coordinates on the affine Lie group $\hat G$ to construct
a new first-order differential
representation of affine $g\times g$, and the corresponding second-order
representation of the (left- and right-mover) affine-Sugawara constructions.

An outline of the paper is as follows.
Part \ref{part1} begins with a review of the new representation of
affine $g\times g$.
We go on to discuss the representation in further detail,
including the construction of the coordinate-space representation of
the primary states of affine $g\times g$.
In Part \ref{part2}, we use
the new representation of the currents, and the corresponding WZW Hamiltonian,
to find  two new equivalent action formulations of
WZW theory in terms of the elements $\hg\in\hat G$ of the affine Lie group.
In these formulations, the target space is the affine Lie group,
\begin{equation}
\hg:\quad {\cal B}\mapsto \hat G
\end{equation}
 while the base space ${\cal B}$ may be either one-dimensional (called
the mechanical formulation on $\hat G$) or two-dimensional (called the
field theory on $\hat G$).
In Part \ref{part3}, we develop a translation dictionary in which the
new formulations on $\hat G$ are understood as mode formulations
of the conventional WZW formulation on the Lie group $G$.
Using the translation dictionary, we also find a three-dimensional
formulation of WZW theory on the Lie group.

For comparison with our results, we first recall the conventional
formulation \cite{Nov,Wit} of WZW theory on $G$,
\begin{subequations}
\begin{equation}
L_{WZW}={k\over8\p}\int d\s \left[
\et_{ab}e_i{}^a e_j{}^b (\pt x^i\pt x^j - \ps x^i \ps x^j)
 + 2B_{ij}\pt x^i \ps x^j \right]
\label{firsteq}
\end{equation}
\begin{equation}
i,a=1\ldots\dim g
\label{wzwaction}
\end{equation}
\begin{equation}
S_{WZW}=-{k\over2\p\chi}\int d\t d\s \Tr(g^{-1}\pa g g^{-1}\bpa g)
-{k\over12\p\chi}\int \Tr(g^{-1}dg)^3
\label{firstg}
\end{equation}
\end{subequations}
where $k$ is the level of the affine algebra.
In the sigma model form (\ref{firsteq}) of the Lagrangian,
$x^i(\t,\s)$, $e_i{}^a(x(\t,\s))$ and $B_{ij}(x(\t,\s))$
are respectively
the coordinates on $G$, the vielbein on $G$, and the antisymmetric tensor
field on $G$.
In the $g\in G$ form (\ref{firstg}) of the action,
$\chi$ is a trace normalization and $\pa=(\pt+\ps)/2$,
$\bpa=(\pt-\ps)/2$.

Our first new formulation of WZW theory
on the affine Lie group $\hat G$ is a mechanical system with
Lagrangian
\begin{subequations}
\begin{equation}
L_M={k\over4}\et_{ab}e_{i\m}{}^{a,-m}e_{j\n}{}^{bm} \pt x^{i\m}\pt x^{j\n}
  - {k\over4} \et^{ab} \hO_{a,-m}{}^\ys \hO_{bm}{}^\ys  %   \nonumber \\
  + k \pt x^{i\m} \left(e_{i\m}{}^\ys + {1\over2} e_{i\m}{}^{am}\hO_{am}{}^\ys
   \right)
\end{equation}
\begin{equation}
={k\over4}\et_{ab}e_{i\m}{}^{a,-m}e_{j\n}{}^{bm} \pt x^{i\m}\pt x^{j\n}
  - {k\over4} \left( \hO_{am}{}^\ys+2\hB_{i\m,j\n}
    \pt x^{j\n}e_{am}{}^{i\m}\right)\et^{ab}\hO_{b,-m}{}^\ys
    \hspace*{21pt}
\label{Bform0}
\end{equation}
\begin{equation}
\vphantom{k\over4}
i,a=1\ldots\dim g, \qquad \m,m\in\Z
\end{equation}
\end{subequations}
where $x^{i\m}(\t)$ are the coordinates on $\hat G$ and
$e(x(\t))$ and $\hO(x(\t))$ are the
vielbein on $\hat G$ and the adjoint action of the affine group
element $\hg$.  The index $\ys$ (on the vielbein and
the adjoint action) labels the extra dimension
of the affine group manifold
corresponding to the central term in the affine algebra.
In (\ref{Bform0}), $\hB_{i\m,j\n}(x(\t))$
is an antisymmetric tensor field on the affine
group, whose role in the mechanical system
is analogous to that of the conventional antisymmetric
tensor field $B_{ij}(x(\t,\s))$ on the Lie group.

The group element form of this mechanical action is
\begin{equation}
S_M=-{k\over4\chi}\int d\t \hTr(\hg^{-1}\pt\hg\hg^{-1}\pt\hg
 -\hg^{-1}\hg'\hg^{-1}\hg')
+{k\over2\chi}\int d\t \int_0^1 d\rho\, \vare^{AB}\hTr(\hg^{-1}\hg'\hg^{-1}
 \pa_{A}\hg\hg^{-1}\pa_{B}\hg)
\end{equation}
where $\hg\in\hat G$. The rescaled trace $\hTr$ and
the symbol $\hg'$ (which is closely related to $\hg$) are defined in the text.
This form of the action exhibits a two-dimensional WZW term
on the affine group.

The actions and Lagrangians above are equal
\begin{equation}
S_{WZW}=S_M, \qquad L_{WZW}=L_M
\end{equation}
under the translation dictionary,
and in fact their kinetic and WZW terms are separately equal.  We note
in particular the various
forms of the WZW term,
$$
-{k\over12\p\chi}\int \Tr(g^{-1}dg)^3 =
  {k\over4\p}\int d\t d\s B_{ij}\pt x^i \ps x^j
= {k\over2\chi}\int d\t \int_0^1 d\rho\, \vare^{AB}\hTr(\hg^{-1}\hg'\hg^{-1}
 \pa_{A}\hg\hg^{-1}\pa_{B}\hg)
$$
\begin{equation}
=   {k\over2} \int d\t \hB_{i\m,j\n}
    \pt x^{i\m}e_{am}{}^{j\n}\et^{ab}\hO_{b,-m}{}^\ys
=  k \int d\t
\pt x^{i\m} \left(e_{i\m}{}^\ys + {1\over2} e_{i\m}{}^{am}\hO_{am}{}^\ys
   \right)
\end{equation}
which now range from three- to one-dimensional.

Our second new formulation of WZW theory
 on $\hat G$ is a constrained two-dimensional field theory,
\begin{eqnarray}
S_F&=&-{k\over2\p\chi}\int d\t d\s
\hat{\Tr}\left(\hg^{-1}\pa\hg
\hg^{-1}\bar\pa\hg
 \right)
 - {k\over 12\p \chi} \int \hTr(\hg^{-1}d\hg)^3
\nonumber \\
& &
 +{k\over\chi}\int d\t d\s
\hTr( \l \hg^{-1}(\ps \hg-\hg'))
\label{traction0}
\end{eqnarray}
with a three-dimensional WZW term on the affine group.  The last term
of (\ref{traction0}) is the constraint term, where $\l$ is the multiplier.
The first two terms of this action, without the
constraint term, were considered as a theory in Ref.~\cite{schwim},
and the theory was found to have an infinite degeneracy
in that case.
In the formulation (\ref{traction0}),
it is the role of the constraint to remove that
degeneracy and to implement the
classical equivalence with the conventional formulation
of WZW theory on the Lie
group.

We also find formal quantum equivalence of the formulations on $\hat G$
with the conventional formulation of WZW on $G$,
\begin{equation}
\int(\cD_M\hg)\, e^{iS_M}=\int(\cD\l\cD_F\hg) \,e^{iS_F}
=\int(\cD g) \,e^{iS_{WZW}}
\label{hooey}
\end{equation}
where the equalities hold up to irrelevant constants.
In (\ref{hooey}), the conventional formal
WZW measure $\cD g$ is a product of Haar
measures on $G$ at each spacetime point.
Similarly, the affine measures $\cD_M\hg$ and $\cD_F\hg$ are
appropriate spacetime products of formal Haar measures
on $\hat G$.

Our final form of WZW theory is a constrained
three-dimensional field theory on
the Lie group
\begin{eqnarray}
S_3&=&-{k\over4\p^2\chi}\int d\t d\s d\ts \Tr \left(g^{-1}\pa g
 g^{-1}\bar\pa g\right) - {k\over 24\p^2 \chi} \int \Tr(g^{-1} dg)^3
\! \wedge \! d\ts
\nonumber \\
& &
 +{k\over2\p\chi}\int d\t d\s d\ts
 \Tr(\l g^{-1}(\ps-\pa_{\ts})g)
\end{eqnarray}
with a four-dimensional WZW term.

The interested reader may profit from an early glance at the picture
(\ref{picture}), which is a schematic presentation of the
relations among all four formulations of WZW theory discussed in this
paper.

\newpage
\part{Affine Lie groups and affine Lie derivatives}
\label{part1}

\section{First order differential representation of affine $g\times g$}
\label{ALA}

In this section, we review the first-order differential representation
of affine Lie algebra recently given by Halpern and Sochen \cite{tor}.

We begin with the current modes $J_a(m)$ of untwisted simple affine Lie $g$
\cite{KM,BH},
\begin{subequations}
\begin{equation}
[J_a(m),J_b(n)]=if_{ab}{}^c J_c(m+n)+mk\et_{ab}\d_{m+n,0}
\end{equation}
\begin{equation}
a,b=1\ldots\dim g, \qquad m,n\in\Z
\end{equation}
\end{subequations}
where $k$ is the level of the affine algebra and
$f_{ab}{}^c$ and $\et_{ab}$ are the structure constants
and Killing metric of Lie $g$.
It is convenient to write the affine algebra as an infinite dimensional
Lie algebra,
\begin{subequations}
\begin{equation}
[\J_L,\J_M]=if_{LM}{}^N\J_N
\label{summation}
\end{equation}
\begin{equation}
\J_L=(J_a(m),k), \qquad L=(am,\ys)
\end{equation}
\begin{equation}
f_{am,bn}{}^{cp}=f_{ab}{}^c \d_{m+n,p}, \qquad
f_{am,bn}{}^\ys=-im\et_{ab} \d_{m+n,0}
\label{structure}
\end{equation}
\end{subequations}
where $L$ is the general tangent-space index.  Here the
central element $k$ is included as a generator, and the
non-zero structure constants $f_{LM}{}^N$ are given in (\ref{structure}).
The summation convention in (\ref{summation}) is generally assumed
throughout this paper.  The adjoint matrix representation of the affine
algebra,
\begin{equation}
(\hT_L^{adj})_M{}^N=-if_{LM}{}^N, \qquad
[\hT_L^{adj},\hT_M^{adj}]=if_{LM}{}^N \hT_N^{adj}
\label{tadj}
\end{equation}
is constructed from the structure constants as usual.

We also introduce the object $\het_{LM}$
\begin{equation}
\het_{am,bn}=\et_{ab}\d_{m+n,0}, \qquad
\het_{\ys,L}=\het_{L,\ys}=0
\label{etahat}
\end{equation}
which will be called the rescaled Killing metric of the affine algebra.
The relation of the rescaled Killing metric to the formal Killing
metric of the affine algebra will be discussed in Section \ref{traces},
which also discusses other formal issues of this nature.
The rescaled Killing metric can be used to obtain the
totally antisymmetric structure constants of the affine algebra,
\begin{subequations}
\begin{equation}
f_{LMN}\equiv f_{LM}{}^P\het_{PN}
\end{equation}
\begin{equation}
f_{am,bn,cp}=f_{abc}\d_{m+n+p,0}
\label{antisym}
\end{equation}
\begin{equation}
(\hT^{adj}_L)_{MN}=-(\hT^{adj}_L)_{NM}
\label{antsc}
\end{equation}
\end{subequations}
whose non-zero components are given in (\ref{antisym}).  The rescaled
Killing metric arises frequently in the development below, although
we often find it
convenient to simplify mode sums by using the
Kronecker delta.

We consider next the affine Lie group $\hat G$ generated by the
affine algebra, whose arbitrary element $\hga\in\hat G$ can be
written
\begin{equation}
\hga(\J,x,y)=e^{iyk} \hg(J,x).
\label{source}
\end{equation}
Here, $y$ and $x^{i\m}, i=1\ldots\dim g, \m \in \Z$ are the
coordinates on the affine group manifold, and $\hg$ is the reduced affine
group element.  For simplicity, we assume that $\hat G$ is simply connected
and we limit ourselves in this paper to the $\b$-family of bases
\begin{equation}
\hg(J,x)=\exp(i\b^{am}(x)J_a(m))
\label{bfamily}
\end{equation}
for the reduced affine group element.
We also assume that the tangent-space coordinates $\b^{am}$ are invertible
functions of $x$
so that $x^{i\m}(\b)$ is well defined.
An example of this
family of bases is the standard basis,
\begin{equation}
\b^{am}(x)=x^{i\m}e_{i\m}{}^{am}(0)
\end{equation}
where $e_{i\m}{}^{am}(0)$ is the left-invariant
vielbein on $\hat G$ (defined below) at the origin.

The left- and
right-invariant vielbeins
$e_\L{}^L(x)$ and $\be_\L{}^L(x)$ on $\hat G$ are defined as follows,
\begin{subequations}
\begin{equation}
e_\L=-i\hga^{-1}\pa_\L\hga=e_\L{}^L\J_L  , \qquad
\be_\L=-i\hga\pa_\L\hga^{-1}=\be_\L{}^L\J_L
\end{equation}
\begin{equation}
L=(am,\ys), \qquad
\L=(i\m,y)
\end{equation}
\label{vbeins}
\end{subequations}
where $\L$ is the general Einstein index.
Note that we have distinguished the Einstein index $y$ from the
tangent space index $\ys$, both of which are associated to the
coordinate $y$.
The inverse vielbeins may be used to construct the
left- and right-invariant affine Lie derivatives $\E$ and $\bar\E$,
{\samepage
\begin{subequations}
\begin{equation}
\E_L=-ie_L{}^\L\pa_\L, \qquad
\bar\E_L=-i\be_L{}^\L\pa_\L
\end{equation}
\begin{equation}
\E_L\hat\g=\hat\g\J_L, \qquad
\bar\E_L\hat\g=-\J_L\hat\g
\label{Eong}
\end{equation}
\begin{equation}
\E_L=(\E_a(m),\E_\ys), \qquad
\bar\E_L=(\bar\E_a(m),\bar\E_\ys).
\end{equation}
\end{subequations}
}
  The relations in (\ref{Eong})
guarantee that the affine Lie derivatives $\E_L$ and
$\bar\E_L$ satisfy two commuting affine algebras with
central elements $\E_\ys$ and
$\bar\E_\ys$.

{}From (\ref{source}) and (\ref{vbeins}), one finds that
the vielbeins satisfy the relations,
\begin{subequations}
\begin{equation}
e_{i\m}=-i\hg^{-1}\pa_{i\m}\hg=e_{i\m}{}^{am}J_a(m)+e_{i\m}{}^\ys k
\label{unbarred2}
\end{equation}
\begin{equation}
\be_{i\m}=-i\hg\pa_{i\m}\hg^{-1}=\be_{i\m}{}^{am}J_a(m)+\be_{i\m}{}^\ys k
\label{barred2}
\end{equation}
\begin{equation}
e_y{}^L=-\be_y{}^L=\d_y{}^L , \qquad
e_{i\m}{}^L, \be_{i\m}{}^L \mbox{ are independent of } y
\label{delta}
\end{equation}
\begin{equation}
e_\ys{}^\L=-\be_\ys{}^\L=\d_\ys{}^\L  , \qquad
e_{am}{}^\L, \be_{am}{}^\L \mbox{ are independent of } y
\end{equation}
\begin{equation}
e_{am}{}^{i\m}e_{i\m}{}^{bn}
=\be_{am}{}^{i\m}\be_{i\m}{}^{bn}
=\d_{am}{}^{bn}  , \qquad
e_{i\m}{}^{am}e_{am}{}^{j\n}=
\be_{i\m}{}^{am}\be_{am}{}^{j\n}=
\d_{i\m}{}^{j\n}
\end{equation}
\begin{equation}
e_{am}{}^y = -e_{am}{}^{i\m}e_{i\m}{}^\ys, \qquad
\be_{am}{}^y=\be_{am}{}^{i\m}\be_{i\m}{}^\ys
\end{equation}
\end{subequations}
and (\ref{delta}) also implies that $\bar\E_\ys=-\E_\ys=i\pa_\ys$.

The induced action of the affine Lie derivatives on
the reduced group element $\hg$ is described by the
reduced affine Lie derivatives $E$ and $\bar E$,
\begin{subequations}
\begin{equation}
E_a(m)=-i e_{am}{}^{i\m}\cD_{i\m}, \qquad
\bE_a(m)=-i \be_{am}{}^{i\m}\bar\cD_{i\m}
\label{redaff}
\end{equation}
\begin{equation}
\cD_{i\m}\equiv\pa_{i\m}-ik e_{i\m}{}^\ys, \qquad
\bar\cD_{i\m}\equiv\pa_{i\m}+ik\be_{i\m}{}^\ys
\label{covar}
\end{equation}
\begin{equation}
E_a(m)\hg=\hg J_a(m),\qquad
\bE_a(m)\hg=-J_a(m)\hg
\label{indact}
\end{equation}
\label{quantE}
\end{subequations}
where $\cD$ and $\bar\cD$ are called the covariant derivatives.
It follows from (\ref{indact}) that
the reduced affine Lie derivatives satisfy two commuting copies of the
affine algebra,
\begin{subequations}
\begin{equation}
[E_a(m),E_b(n)]=if_{ab}{}^c E_c(m+n)+mk\et_{ab}\d_{m+n,0}
\label{Escommute}
\end{equation}
\begin{equation}
[\bE_a(m),\bE_b(n)]=if_{ab}{}^c \bE_c(m+n)-mk\et_{ab}\d_{m+n,0}
\end{equation}
\begin{equation}
[E_a(m),\bE_b(n)]=0
\label{bothcommute}
\end{equation}
\label{Ealgebra}
\end{subequations}
at level $k$ and $-k$ respectively.
Although we will only do so in Section \ref{primst},
one may obtain two commuting copies of the affine algebra at
the same level $k$ by defining $\bE'_a(m)\equiv\bE_a(-m)$.

The reduced affine Lie derivatives
(\ref{redaff}) are a first-order differential representation of
the currents of
affine $g\times g$.  Other first-order differential representations of
affine Lie algebra are known, such as the coadjoint orbit representations
in Refs.~\cite{bf} and \cite{t}, but these provide only a single chiral copy of
the algebra.

Although the construction above guarantees that the
reduced affine Lie derivatives (\ref{quantE})
satisfy the algebra (\ref{Ealgebra}) of affine $g\times g$, it
is useful to have the machinery to check these relations
directly.

We need in particular the Cartan-Maurer and inverse Cartan-Maurer
identities for the left-invariant vielbein,
\begin{subequations}
\begin{equation}
\pa_{\L}e_{\Ga}{}^{L}-\pa_{\Ga}e_{\L}{}^{L}
= e_{\L}{}^{M}e_{\Ga}{}^{N}f_{MN}{}^{L}
\label{CMa}
\end{equation}
\begin{equation}
e_L{}^{\L}\pa_{\L}e_M{}^\Ga-e_M{}^{\L}\pa_{\L}e_L{}^\Ga =
f_{ML}{}^N e_N{}^\Ga
\end{equation}
\label{CM}
\end{subequations}
which follow from (\ref{unbarred2}).  The same relations with $e\ra\bar e$
hold for the right-invariant vielbein.

These identities are sufficient to compute the curvature of the covariant
derivatives in (\ref{covar}),
\begin{equation}
[\cD_{i\m},\cD_{j\n}]=k\sum_n ne_{i\m}{}^{a,-n} \et_{ab} e_{j\n}{}^{bn},
\qquad
[\bar\cD_{i\m},\bar\cD_{j\n}]
=-k\sum_n n\be_{i\m}{}^{a,-n} \et_{ab} \be_{j\n}{}^{bn}
\end{equation}
and also to check that the reduced affine Lie derivatives satisfy
the affine Lie algebras (\ref{Ealgebra}a,b).

To check the commutator of $E$ and $\bE$, we first introduce
the adjoint action $\hO$ of $\hg$, which satisfies
\begin{subequations}
\begin{equation}
\hg\J_L\hg^{-1}=\hO_L{}^M\J_M, \qquad
\hO(x)=\hg^{-1}(\hT^{adj},x), \qquad
\hO_\ys{}^L=\d_\ys{}^L
\label{omdef}
\end{equation}
\begin{equation}
e_{am}{}^{i\m}\pa_{i\m}\hO_L{}^M=-f_{am,L}{}^N\hO_N{}^M, \qquad
\bar e_{am}{}^{i\m}\pa_{i\m}\hO_L{}^M=\hO_L{}^N  f_{am,N}{}^M
\label{Oderiv}
\end{equation}
\begin{equation}
\hO_{am}{}^{cp}\het_{cp,dq}\hO_{bn}{}^{dq} = \het_{am,bn}
\label{orthog}
\end{equation}
\label{omega}
\end{subequations}
where the matrix-valued group element
$\hg(\hT^{adj},x)$ is $\hg(J,x)$ with
$J\rightarrow \hT^{adj}$ (see eq.(\ref{tadj})).
The quantity $\het$ is the rescaled Killing metric (\ref{etahat}), and
the pseudo-orthogonality relation
(\ref{orthog}) follows from the
antisymmetry of $\hT^{adj}$.

We also need the relations between the right- and left-invariant
quantities,
\begin{subequations}
\begin{equation}
\be_\L{}^L=-e_\L{}^M\hO_M{}^L, \qquad
\be_L{}^\L=-(\hO^{-1})_L{}^M e_M{}^\L
\label{erel}
\end{equation}
\begin{equation}
\bE_a(m)=(\hO^{-1})_{am}{}^{bn}(\hO_{bn}{}^\ys k - E_b(n))
\label{Erel}
\end{equation}
\label{leftright}
\end{subequations}
which follow from eqs.(\ref{omdef}) and (\ref{vbeins}).
Using (\ref{Erel}), (\ref{Escommute}), and (\ref{Oderiv}),
it is straightforward to
verify that $\bE$ commutes with $E$.

\section{The primary states of affine $g\times g$}
\label{primst}

The reduced affine Lie derivatives
(\ref{quantE}) are a coordinate-space representation of
affine $g\times g$, so it is natural to consider the
coordinate-space representation of the primary states of affine $g \times g$.

To construct these states, we
begin with a matrix irrep $T$ of the Lie algebra $g$,
\begin{equation}
[T_a,T_b]=if_{ab}{}^c T_c
\end{equation}
and introduce the corresponding chiral affine primary states $|R(T)\rangle$,
which satisfy
\begin{subequations}
\begin{equation}
J_a(m\ge0)|R(T)\rangle^I = \d_{m,0}|R(T)\rangle^J (T_a)_J{}^I
\end{equation}
\begin{equation}
{}_J\langle R(\bar T)|J_a(m\le0) = (T_a)_J{}^K\, {}_K\langle R(\bar T)|
  \d_{m,0}
\end{equation}
\begin{equation}
{}_J\langle R(\bar T)|R(T)\rangle^I = \d_{J}{}^I, \qquad
I,J,K=1\ldots\dim T.
\end{equation}
\end{subequations}
Then, the primary states $\psi(T,x)$
of affine $g\times g$ are constructed as
\begin{equation}
\psi(T,x)_J{}^I \equiv {}_J\langle R(\bar T)|\hat g(J,x) | R(T)\rangle^I
= {}_J\langle R(\bar T)|e^{i\b^{am}(x)J_a(m)} | R(T)\rangle^I.
\end{equation}
Using the induced action (\ref{indact}) of the reduced affine Lie
derivatives,
it is easily checked that these states are primary,
\begin{subequations}
\begin{equation}
E_a(m\ge0)\psi(T,x)_J{}^I=\d_{m,0}\psi(T,x)_J{}^K(T_a)_K{}^I
\end{equation}
\begin{equation}
\bar E_a'(m\ge0)\psi(T,x)_J{}^I=-\d_{m,0}(T_a)_J{}^K\psi(T,x)_K{}^I
\end{equation}
\end{subequations}
where $\bE_a'(m)=\bE_a(-m)$.
The other states in the modules of affine $g \times g$ are
constructed as usual by the action of the negative modes
$E_a(m<0)$ and $\bar E_a'(m<0)$ on the primary states.

We remark that
the primary states may be expanded about the origin to any desired
order
\begin{equation}
\psi(T,x)=1+i\b^{a0}T_a-{1\over2}(\b^{a0}T_a)^2-
{1\over2}\sum_{m=1}^\infty \b^{am}\b^{b,-m}(if_{ab}{}^cT_c+km\et_{ab})
+\cO(\b^3)
\end{equation}
in powers of the tangent-space coordinates $\b^{am}(x)$.
This expansion is closely related to a high-level expansion of the
primary states,
\begin{equation}
\b^{am}={y^{am}\over k}, \qquad
\psi(T,x)=1+{1\over k}\left(i y^{a0}T_a-{1\over2}\sum_{m=1}^\infty
y^{am}y^{b,-m}m\et_{ab}\right) + \cO(k^{-2})
\end{equation}
whose leading terms
correspond to an abelian contraction \cite{Review} of the
affine algebra.
A different high-level expansion of the primary fields is
\begin{equation}
\b^{a,m\ne0}={z^{am}\over k}, \qquad
\psi(T,x)=g(T,x)(1+\cO(k^{-1}))
\end{equation}
where $g(T,x)=\exp(i\b^{a0}(x)T_a)\in G$ is the Lie group element.
This expansion corresponds to another contraction \cite{Review} of the
affine algebra, in which only the non-zero modes are abelian.

\section{The antisymmetric tensor field on the affine group}

In this section, we find
an antisymmetric tensor field $\hB_{i\m,j\n}=-\hB_{j\n,i\m}$ on the
affine Lie group $\hat G$, which, as we shall see, is analogous to
the antisymmetric tensor field $B_{ij}=-B_{ji}$ of the
conventional WZW model on $G$.

We begin by choosing a $\b$-basis
\begin{equation}
\hg(J,x)=\exp\left(i\b^{am}(x)J_a(m)\right)
\label{basis}
\end{equation}
for the reduced affine group element $\hg$.
In any such basis, the vielbein has the explicit form,
\begin{subequations}
\begin{equation}
e_{i\m}{}^L(x)=\pa_{i\m}\b^{am}(x)M(x)_{am}{}^L, \qquad
e_y{}^L=\d_\ys{}^L
\end{equation}
\begin{equation}
M(x)\equiv {1-\hO^{-1} \over \log\hO}
\end{equation}
\label{viel1}
\end{subequations}
where $\hO$ is the adjoint action in (\ref{omega}).  This form
generalizes the result given
for the standard basis
$\b^{am}=x^{i\m}e_{i\m}{}^{am}(0)$ in Ref.~\cite{tor}.
We may eliminate $\b^{am}$ in (\ref{viel1}) to find
the relation between $e^\ys$ and $\hO^\ys$,
\begin{subequations}
\begin{equation}
e_{i\m}{}^\ys={1\over2}\left(\hB_{i\m,j\n}e_{a,-m}{}^{j\n}\et^{ab}
             -e_{i\m}{}^{bm}\right)\hO_{bm}{}^\ys
\label{ey}
\end{equation}
\begin{equation}
\hB_{i\m,j\n}\equiv e_{i\m}{}^{am} \N_{am}{}^{bn}
\het_{bn,cp} e_{j\n}{}^{cp}
\label{Banti}
\end{equation}
\begin{equation}
\N(x) \equiv
  {(\hO^{-1}-\hO) + 2\log \hO \over
   (\hO-1)(\hO^{-1}-1) }
\end{equation}
\label{Bdef}
\end{subequations}
where $\et^{ab}$ is the inverse Killing metric of Lie $g$ and
$\hB_{i\m,j\n}$ is the desired tensor field on $\hat G$.

We know that $\hO$ in (\ref{orthog}) is pseudo-orthogonal,
so the matrix
$\N\het$ in (\ref{Banti}) is antisymmetric.  It follows that
$\hB_{i\m,j\n}$ is antisymmetric,
\begin{equation}
\hB_{i\m,j\n}=-\hB_{j\n,i\m}.
\end{equation}
A useful property of the antisymmetric tensor field is
\begin{equation}
\pa_{i\m}(\hB_{j\n,k\r}e_{a,-m}{}^{k\r}\et^{ab}\hO_{bm}{}^\ys) -
\left( i\m \lra j\n \right) =
e_{i\m}{}^{bn}e_{j\n}{}^{am}f_{am,bn}{}^{cp}\hO_{cp}{}^\ys
\label{Bident}
\end{equation}
which follows from (\ref{ey}) and the Cartan-Maurer identity (\ref{CMa}) for
$e_{i\m}{}^\ys$.

With (\ref{erel}) and (\ref{ey}), we may rewrite the reduced affine
Lie derivatives (\ref{quantE}) in the $\hB$-form,
\begin{subequations}
\begin{equation}
E_a(m)=-ie_{am}{}^{i\m}\cD_{i\m}(\hB)
+{1\over2}k\hO_{am}{}^\ys, \qquad
\bE_a(m)=-i\be_{am}{}^{i\m}\cD_{i\m}(\hB)
-{1\over2}k\hO_{am}{}^\ys
\end{equation}
\begin{equation}
\cD_{i\m}(\hB)\equiv
\pa_{i\m}-{i\over2}k\hB_{i\m,j\n}e_{a,-m}{}^{j\n}\et^{ab}\hO_{bm}{}^\ys.
\end{equation}
\label{quantE2}
\end{subequations}
Using the identity
(\ref{Bident}) and the steps of the previous section,
it is straightforward
to check explicitly that the $\hB$-form  of the
reduced affine Lie derivatives
satisfies the algebra (\ref{Ealgebra}) of affine $g\times g$.

In Section \ref{traces},
we will find that $\hB_{i\m,j\n}$ satisfies other identities which
show further analogy with the conventional WZW tensor field $B_{ij}$ on
$G$ (see also Section \ref{modestolocal}).
Further discussion of the operator currents (\ref{quantE2}) is given in
Section \ref{lflc}.

\section{Bracket representation of affine $g\times g$}
\label{brackrep}

To construct classical dynamics on the affine Lie group,
we need the
Poisson bracket representation which corresponds to the
first-order differential representation
(\ref{quantE}) or (\ref{quantE2}) of affine $g \times g$.
The bracket representation may be
obtained by the usual prescription,
\begin{subequations}
\begin{equation}
\pa_{i\m}\rightarrow ip_{i\m}
\label{substitution}
\end{equation}
\begin{equation}
\{x^{i\m},p_{j\n}\} = i\d_{j\n}{}^{i\m}, \qquad
\{x^{i\m},x^{j\n}\} = \{p_{i\m},p_{j\n}\} = 0
\label{pb}
\end{equation}
\end{subequations}
where $\{A,B\}$ is Poisson bracket and
$p_{i\m}$ are (classical) canonical
momenta.

Using the substitution (\ref{substitution}) in the reduced affine
Lie derivatives (\ref{quantE}), we
find one form of the classical current modes
\begin{subequations}
\begin{equation}
E_a(m)= e_{am}{}^{i\m} p_{i\m}^- , \qquad
\bE_a(m)= \be_{am}{}^{i\m} p_{i\m}^+
\end{equation}
\begin{equation}
p_{i\m}^-\equiv p_{i\m}-ke_{i\m}{}^\ys, \qquad
p_{i\m}^+\equiv p_{i\m}+k\be_{i\m}{}^\ys.
\label{pminus}
\end{equation}
\label{classE}
\end{subequations}
Similarly, the equivalent $\hB$-form of the classical current modes,
{\samepage
\begin{subequations}
\begin{equation}
E_a(m)= e_{am}{}^{i\m} p_{i\m}(\hB) + {1\over2}k\hO_{am}{}^\ys, \qquad
\bE_a(m)=\be_{am}{}^{i\m} p_{i\m}(\hB) - {1\over2}k\hO_{am}{}^\ys
\end{equation}
\begin{equation}
p_{i\m}(\hB)\equiv p_{i\m} -
{1\over2}k\hB_{i\m,j\n}e_{a,-m}{}^{j\n}\et^{ab}\hO_{bm}{}^\ys
\label{phat}
\end{equation}
\label{classE2}
\end{subequations}
}
is obtained by the same substitution in the operator $\hB$-form
(\ref{quantE2}).
Following the steps of Section \ref{ALA},
the bracket algebra of affine $g\times g$,
\begin{subequations}
\begin{equation}
\{E_a(m),E_b(n)\}=if_{ab}{}^c E_c(m+n)+mk\et_{ab}\d_{m+n,0}
\end{equation}
\begin{equation}
\{\bE_a(m),\bE_b(n)\}=if_{ab}{}^c \bE_c(m+n)-mk\et_{ab}\d_{m+n,0}
\end{equation}
\begin{equation}
\{E_a(m),\bE_b(n)\}=0
\end{equation}
\label{cEalgebra}
\end{subequations}
is easily verified for both forms of the classical currents.

In what follows, we interchangeably use the terms
classical affine Lie derivatives or classical currents
to refer to $E_a(m)$ and $\bar E_a(m)$
in (\ref{classE},3).

\part{Actions on the affine Lie group $\hat G$}
\label{part2}

\section{WZW as a mechanical system on the affine group}
\label{wzwclass}

In this section we construct a natural action for classical mechanics
on the affine Lie group $\hat G$.  As is clear from its construction,
this mechanical action must be equivalent to the
WZW action on the corresponding Lie group $G$.
The equivalence is studied explicitly in Part \ref{part3}.

The operator WZW Hamiltonian $H=L_g^{ab}\xx J_a J_b+\bar J_a\bar J_b\xx{}_0$
sums the zero modes of left- and right-mover affine-Sugawara
constructions.  When written in terms of the reduced affine Lie
derivatives (\ref{quantE}), the coordinate-space form
$H=L_g^{ab}\xx E_aE_b+\bE_a\bE_b\xx{}_0$ of this Hamiltonian is a
natural Laplacian \cite{tor} on the affine Lie group.

To construct the corresponding action on the affine Lie group,
we begin with the standard classical WZW Hamiltonian
written in terms of
the classical affine Lie derivatives (\ref{classE},3),
\begin{subequations}
\begin{eqnarray}
H&=&{1\over 2k}\het^{am,bn}\left(E_a(m)E_b(n)+\bE_a(m)\bE_b(n)\right)
\label{firstH}
\\
&=&{1\over 2k}\et^{ab}\sum_m\left(E_a(-m)E_b(m)+\bE_a(-m)\bE_b(m)\right)
\end{eqnarray}
\begin{equation}
\het^{am,bn}\equiv\et^{ab}\d_{m+n,0}
\label{hetinv}
\end{equation}
\begin{equation}
\pa_\t A=i\{H,A\}
\label{Heom}
\end{equation}
\end{subequations}
where $\het^{am,bn}$ in (\ref{hetinv}) is the inverse of the rescaled
Killing metric $\het_{am,bn}$ in (\ref{etahat}).  As seen in (\ref{firstH}),
this Hamiltonian
is the natural generalization to affine $g \times g$ of the
 Casimir operator on Lie $g$.

The time dependence of the classical currents
\begin{subequations}
\begin{equation}
\pt E_a(m,\t)=-im E_a(m,\t), \qquad
\pt \bE_a(m,\t)=im \bE_a(m,\t)
\end{equation}
\begin{equation}
E_a(m,\t)=e^{-im\t}E_a(m), \qquad
\bar E_a(m,\t)=e^{im\t}\bar E_a(m)
\label{timeEb}
\end{equation}
\label{timeE}
\end{subequations}
follows immediately from (\ref{Heom}) and the
current algebra (\ref{cEalgebra}).

Using the explicit form (\ref{classE}) of the classical affine Lie
derivatives, we obtain an explicit form of the Hamiltonian
\begin{equation}
H= {1\over k}\et^{ab} e_{a,-m}{}^{i\m}(x)e_{bm}{}^{j\n}(x)p^-_{i\m}p^-_{j\n} +
    \left( {k\over 2}\hO_{a,-m}{}^\ys(x) -
    e_{a,-m}{}^{i\m}(x)p^-_{i\m}  \right) \et^{ab}\hO_{bm}{}^\ys(x) \quad
\label{p-ham}
\end{equation}
where $p^-_{i\m}$ is defined in (\ref{pminus}).
This Hamiltonian describes a classical mechanics whose coordinates $x$
are coordinates on the affine Lie group.
The equivalent $\hB$-form
of the Hamiltonian is
\begin{equation}
H = {1\over k}\et^{ab} e_{a,-m}{}^{i\m}e_{bm}{}^{j\n}p_{i\m}(\hB)p_{j\n}(\hB) +
{k\over 4}\et^{ab} \hO_{a,-m}{}^\ys \hO_{b,m}{}^\ys
\label{Bham}
\end{equation}
where we have suppressed the
$x$-dependence of all quantities,
 and $p_{i\m}(\hB)$ is defined in (\ref{phat}).
The Hamiltonian equations of motion
\begin{subequations}
\begin{eqnarray}
\pt x^{i\m}&=&{2\over k}\et^{ab} e_{a,-m}{}^{i\m}e_{bm}{}^{j\n}p^-_{j\n}
-\et^{ab}e_{a,-m}{}^{i\m}\hO_{bm}{}^\ys \\
&=&{2\over k}\et^{ab} e_{a,-m}{}^{i\m}e_{bm}{}^{j\n}p_{j\n}(\hB)
\end{eqnarray}
\begin{equation}
\pt p_{i\m}=-\pa_{i\m} H
\end{equation}
\label{dots}
\end{subequations}
follow from eqs.(\ref{p-ham},4).

\vskip .2cm
\noindent\underline{Coordinate space}

We turn now to the coordinate-space formulation of the mechanical
theory on $\hat G$.
To begin, one uses eq.(\ref{classE}) or (\ref{classE2}) and
the equations of motion (\ref{dots}) to obtain
\begin{equation}
E_a(m)={k\over2}(\et_{ab}e_{i\m}{}^{b,-m}\pt x^{i\m}+\hO_{am}{}^\ys), \qquad
\bar E_a(m)={k\over2}(\et_{ab}\bar e_{i\m}{}^{b,-m}\pt x^{i\m}-\hO_{am}{}^\ys)
\end{equation}
for the coordinate-space form of the currents.

Similarly, the mechanical action on the affine Lie group
\begin{subequations}
\begin{equation}
S_M=\int d\t L_M, \qquad
L_M= \pt x^{i\m}p_{i\m} - H
\end{equation}
\begin{equation}
L_M={k\over4}\et_{ab}e_{i\m}{}^{a,-m}e_{j\n}{}^{bm} \pt x^{i\m}\pt x^{j\n}
  - {k\over4} \et^{ab} \hO_{a,-m}{}^\ys \hO_{bm}{}^\ys  %   \nonumber \\
  + k \pt x^{i\m} \left(e_{i\m}{}^\ys + {1\over2} e_{i\m}{}^{am}\hO_{am}{}^\ys
   \right)
\end{equation}
\label{myaction}
\end{subequations}
is obtained from the Hamiltonian (\ref{p-ham}) in the usual way.
The equivalent $\hB$-form of the Lagrangian is
\begin{equation}
L_M={k\over4}\et_{ab}e_{i\m}{}^{a,-m}e_{j\n}{}^{bm} \pt x^{i\m}\pt x^{j\n}
  - {k\over4} \left( \hO_{am}{}^\ys+2\hB_{i\m,j\n}
    \pt x^{j\n}e_{am}{}^{i\m}\right)\et^{ab}\hO_{b,-m}{}^\ys.
\label{Baction}
\end{equation}
It is straightforward to check that the associated Lagrange equations of motion
of either form of the action
reproduce the Hamiltonian equations of motion (\ref{dots}).
A third form of the mechanical action, in terms of the reduced affine
group element $\hat g$, is given in Section \ref{mechg}.

This action is one of the central results of this paper.
Although the classical mechanics (\ref{myaction}) or
(\ref{Baction}) on $\hat G$ shows no spatial coordinate $\s$, this formulation
must be equivalent to the conventional
WZW model on $G$ because the Hamiltonians of the two formulations are
isomorphic.  This equivalence is studied in detail in Part \ref{part3}, where
we will see that this action is in
fact a mode formulation of WZW.

\section{WZW as a field theory on the affine group}
\label{sectt}

In the last section, we expressed the WZW model
as a mechanical system on the affine Lie group,
\begin{equation}
\hg(x(\t)):\quad \R \mapsto \hat G .
\end{equation}
  In this
section, we show that the model can also be expressed as a two-dimensional
field theory
\begin{equation}
\hg(x(\t,\s)):\quad \R\times S^1 \mapsto \hat G
\end{equation}
on the affine Lie group.

Mathematically, our task is to extend the base space from
the line $(\t)$ to the cylinder $(\t,\s)$,
thereby promoting the mechanical variables $x^{i\m}(\t)$ to
$\s$-dependent fields $x^{i\m}(\t,\s)$.
The reason that we can make such an equivalent field-theoretic formulation
is that the Hamiltonian of the system
\begin{subequations}
\begin{eqnarray}
H&=&{1\over 2k}\et^{ab}\sum_m\left(E_a(-m)E_b(m)+\bE_a(-m)\bE_b(m)\right) \\
&=& {1\over k}\et^{ab} e_{a,-m}{}^{i\m}e_{bm}{}^{j\n}p_{i\m}(\hB)p_{j\n}(\hB) +
{k\over 4}\et^{ab} \hO_{a,-m}{}^\ys \hO_{b,m}{}^\ys
\end{eqnarray}
\begin{equation}
\pa_\t A=i\{H,A\}
\end{equation}
\label{firstHam}
\end{subequations}
admits a commuting quantity $P$
\begin{subequations}
\begin{eqnarray}
P&=&{1\over 2k}\et^{ab}\sum_m\left(E_a(-m)E_b(m)-\bE_a(-m)\bE_b(m)\right) \\
&=& \hO_{a,-m}{}^\ys \et^{ab}  e_{bm}{}^{i\m} p_{i\m}
\end{eqnarray}
\begin{equation}
\{H,P\}=0
\label{HP}
\end{equation}
\begin{equation}
\pa_\s A\equiv i\{P,A\}
\label{psdef}
\end{equation}
\label{Momentum}
\end{subequations}
which we may interpret, according to (\ref{psdef}),
 as the generator of spatial translations.  This form
of $P$
is the standard WZW momentum, written in terms of the classical
 affine Lie derivatives (\ref{classE2}), and eq.(\ref{HP}) follows
immediately from the current algebra (\ref{cEalgebra}).

The spacetime dependence of any observable $A$ is determined
by this system,
and, in particular, the relations
\begin{equation}
\ps H=\pt P =\pt H = \ps P = 0
\end{equation}
follow from eq.(\ref{HP}).

In this framework, all fields are now $\s$-dependent, and
the bracket relations of the earlier sections,
e.g.
\begin{subequations}
\begin{equation}
\{x^{i\m}(\s),p_{j\n}(\s)\} = i\d_{j\n}{}^{i\m}, \qquad
\{x^{i\m}(\s),x^{j\n}(\s)\}=\{p_{i\m}(\s),p_{j\n}(\s)\}=0
\label{spb}
\end{equation}
\begin{equation}
\{E_a(m,\s),E_b(n,\s)\}=if_{ab}{}^c E_c(m+n,\s)+mk\et_{ab}\d_{m+n,0}
\end{equation}
\end{subequations}
should be read
at equal $\s$.  Moreover, all field products
should be read at equal $\s$, for example
\begin{subequations}
\begin{eqnarray}
P&=&{1\over 2k}\et^{ab}\sum_m\left(E_a(-m,\s)E_b(m,\s)-
   \bE_a(-m,\s)\bE_b(m,\s)\right) \\
&=& \hO_{a,-m}{}^\ys(x(\s)) \et^{ab}  e_{bm}{}^{i\m}(x(\s)) p_{i\m}(\s)
\end{eqnarray}
\end{subequations}
and similarly for the Hamiltonian.

The spatial dependence of the affine Lie derivatives follows immediately
from the current algebra,
\begin{subequations}
\begin{equation}
\ps E_a(m,\t,\s)=-imE_a(m,\t,\s), \qquad
\ps \bE_a(m,\t,\s)=-im\bE_a(m,\t,\s)
\end{equation}
\begin{equation}
E_a(m,\t,\s)=e^{-im\s}E_a(m,\t), \qquad
\bar E_a(m,\t,\s)=e^{-im\s}\bar E_a(m,\t).
\end{equation}
\end{subequations}
Combining this with the known time dependence (\ref{timeE}), we find that
the affine Lie derivatives are chiral,
\begin{equation}
E_a(m,\t,\s)=e^{-im(\t+\s)}E_a(m), \qquad
\bar E_a(m,\t,\s)=e^{im(\t-\s)}\bar E_a(m).
\end{equation}
The usual local chiral currents are then identified as
\begin{subequations}
\begin{equation}
E_a(\t,\s)\equiv\sum_m E_a(m,\t,\s)=\sum_m e^{-im(\t+\s)}E_a(m)
\end{equation}
\begin{equation}
\bE_a(\t,\s)\equiv\sum_m \bE_a(m,\t,\s)=\sum_m e^{im(\t-\s)}\bE_a(m)
\end{equation}
\label{locE}
\end{subequations}
and the usual local current algebra
\begin{subequations}
\begin{equation}
\{E_a(\t,\s),E_b(\t,\s')\}=2\p i[f_{ab}{}^c E_c(\t,\s) \d(\s-\s')
 + k \et_{ab}\ps\d(\s-\s')]
\end{equation}
\begin{equation}
\{\bE_a(\t,\s),\bE_b(\t,\s')\}=2\p i[f_{ab}{}^c \bE_c(\t,\s) \d(\s-\s')
 - k \et_{ab}\ps\d(\s-\s')]
\end{equation}
\begin{equation}
\{E_a(\t,\s),\bE_b(\t,\s')\}=0
\end{equation}
\label{localE}
\end{subequations}
follows from the
mode algebra (\ref{cEalgebra}).
Further discussion of these local currents is given in
Section \ref{lflc}.

The spacetime derivatives of the canonical variables,
\begin{subequations}
\begin{equation}
\pt x^{i\m}={2\over k}\et^{ab} e_{a,-m}{}^{i\m}e_{bm}{}^{j\n}p_{j\n}(\hB)
\label{xdot}
\end{equation}
\begin{equation}
\ps x^{i\m}=\et^{ab}e_{a,-m}{}^{i\m}\hO_{bm}{}^\ys
\label{xprime}
\end{equation}
\begin{equation}
\pt p_{i\m}=-\pa_{i\m} H
\end{equation}
\begin{equation}
\ps p_{i\m}=-p_{j\n}\pa_{i\m}(e_{a,-m}{}^{j\n}\et^{ab}\hO_{bm}{}^\ys)
\label{pprime}
\end{equation}
\label{dotprime}
\end{subequations}
also
follow from the equal-$\s$ brackets (\ref{spb}) and the explicit forms of
$H$ and $P$.

\vskip .2cm \noindent
\underline{Spatial constraints}

The procedure described above to extend the base space via commuting
$P$ operators is quite general.
One may guarantee that the extended
theory (on the extended base space) is equivalent
to the theory on the unextended base space by considering the spatial
derivative relations (such as (\ref{xprime},d)) to be constraints on
the canonical variables.
In a functional formulation, this prescription corresponds to the
functional identity
{\samepage
\begin{subequations}
\begin{equation}
\int\left(\prod_\t dx(\t)\right)F[x(\t)]=
\int\left(\prod_{\t,\s}dx(\t,\s)\right)
\det\left(\d{\cal C}\over\d x\right) F[x(\t,\s)]
\d[{\cal C}(x(\t,\s))]
\label{measdelt}
\end{equation}
\begin{equation}
{\cal C}\equiv \ps x(\t,\s)-i\{P,x(\t,\s)\}
\end{equation}
\label{meas1}
\end{subequations}
}
and similarly for the canonical momenta.
The functional delta function in (\ref{measdelt}) enforces the spatial
constraints in the extended theory, and
it follows that the spatial
constraints can be implemented as usual in a Dirac formulation of the theory.

In the case at hand, this prescription guarantees equivalence of the
field-theoretic formulation on $\hat G$ with the mechanical formulation
(\ref{Baction}) on $\hat G$, and hence with the conventional WZW
model itself.

To be more explicit about the equivalence with the mechanical
formulation, we first rewrite the constraint (\ref{xprime}) in terms
of the tangent-space fields $\b^{am}$,
\begin{subequations}
\begin{eqnarray}
\ps\b^{am}-im\b^{am}&=&(\ps x^{i\m}-\et^{bc}e_{b,-n}{}^{i\m}\hO_{cn}{}^\ys)
\pa_{i\m}\b^{am}
\label{yup}\\
&=& 0
\end{eqnarray}
\begin{equation}
\b^{am}(x(\t,\s))=e^{im\s}\b^{am}(x(\t))
\label{bdep}
\end{equation}
\label{bconst}
\end{subequations}
whose simple $\s$-dependence is given in (\ref{bdep}).
The identity (\ref{yup})
is obtained by using eq.(\ref{xprime}) and the explicit forms
of the vielbein and the adjoint action given in Appendix B\@.
It follows by chain rule from (\ref{yup}) that the measure factor
in (\ref{measdelt}) is effectively constant
\begin{equation}
\det\left(\d\C\over\d x\right)=\mbox{field-independent}
\label{meas2}
\end{equation}
and can be ignored.  It also follows from (\ref{bdep})
that averages $\langle~\rangle$
in the mechanical theory ($\hat G_M$) and the
field theory ($\hat G_F$) on $\hat G$ are simply related,
so long as the
formal functional measures (see Section \ref{Funmeas})
of $\hat G_M$ and $\hat G_F$ are suitably
adjusted.  Specifically, one has
\begin{equation}
\langle \F[\b^{am}(x(\t,\s))]\rangle_{\hat G_F} =
\langle \F[\b^{am}(x(\t))e^{im\s}]\rangle_{\hat G_M}
\label{hghgtrans}
\end{equation}
for any  function $\F$, which includes averages over products
of  the affine group elements.

There is another formulation of the theory, on the constrained subspace,
in which the spatial derivative relations are identities (as in
conventional formulations).
As a first step in this formulation, use the constraint (\ref{xprime})
to reexpress the
Hamiltonian and the momentum
\begin{subequations}
\begin{equation}
H={1\over k}\et^{ab}e_{a,-m}{}^{i\m}e_{bm}{}^{j\n}p_{i\m}(\hB)p_{j\n}(\hB) +
  {k\over 4}\et_{ab}e_{i\m}{}^{a,-m}e_{j\n}{}^{bm}\ps x^{i\m}\ps x^{j\n}
\label{xpham}
\end{equation}
\begin{equation}
P= p_{i\m}\ps x^{i\m}
\end{equation}
\begin{equation}
p_{i\m}(\hB)=p_{i\m}-{k\over 2} \hB_{i\m,j\n}\pa_\s x^{j\n}
\end{equation}
\label{newHP}
\end{subequations}
in terms of $\ps x$.
This form of the system is complete with the canonical brackets (\ref{spb})
and the constraints (\ref{xprime},d), or with the canonical brackets and an
auxiliary set of equal-$\s$ brackets which must be computed from the
constraints.  As examples of the auxiliary set, we have
\begin{subequations}
\begin{equation}
\{ x^{i\m}(\s),\ps x^{j\n}(\s)\}  =
\{ x^{i\m}(\s), \et^{ab}e_{a,-m}{}^{i\m}(x(\s))\hO_{bm}{}^\ys(x(\s))\} = 0
 \end{equation}
 \begin{eqnarray}
\{ p_{i\m}(\s),\ps x^{j\n}(\s)\} &=&
\{ p_{i\m}(\s), \et^{ab}e_{a,-m}{}^{j\n}(x(\s))\hO_{bm}{}^\ys(x(\s))\}
 \nonumber \\
&=&
-i \et^{ab}\pa_{i\m}(e_{a,-m}{}^{j\n}(x(\s))\hO_{bm}{}^\ys(x(\s)))
\qquad \qquad
 \end{eqnarray}
 \begin{eqnarray}
\{ p_{i\m}(\s),\ps p_{j\n}(\s)\} &=&
-\{ p_{i\m}(\s), p_{k\rho}(\s)\pa_{j\n}(e_{a,-m}{}^{k\rho}(x(\s))\et^{ab}
\hO_{bm}{}^\ys(x(\s)))\} \nonumber \\ &=&
i p_{k\rho}(\s)\pa_{i\m}\pa_{j\n}(e_{a,-m}{}^{k\rho}(x(\s))\et^{ab}
\hO_{bm}{}^\ys(x(\s))).
\end{eqnarray}
\label{extend}
\end{subequations}
Using these auxiliary brackets, one now obtains the conventional identity
\begin{equation}
\ps x^{i\m}\equiv i\{P,x^{i\m}\} = i\ps x^{j\n}\{p_{j\n},x^{i\m}\}=
 \ps x^{i\m}
\end{equation}
as expected on the constrained subspace.

\vskip .2cm
\noindent \underline{Density formulation}

The Hamiltonian systems above are unconventional formulations of a field
theory because they are not written in terms of spatial densities.
Since $H$ and $P$ are independent of $\s$, however, we can define
the densities as proportional to the Hamiltonian and momentum,
\begin{subequations}
\begin{equation}
\H\equiv{H\over2\p}, \qquad
H=\int d\s \H
\end{equation}
\begin{equation}
{\cal P}\equiv{P\over2\p}, \qquad
P=\int d\s {\cal P}.
\end{equation}
\end{subequations}
These $\s$-independent
densities may be used for either dynamical system (\ref{firstHam},4)
or (\ref{newHP}), but the constraints (or the auxiliary brackets)
must be included in the latter case.
Because the densities are $\s$-independent,
 one has the bracket equations of motion
\begin{eqnarray}
\pt A(\s)&=&i\{H,A(\s)\}=\int d\s' \{\H(\s'),A(\s)\}=
i\int d\s' \{\H(\s),A(\s)\} \nonumber \\
&=& 2\p i \{\H(\s),A(\s)\}
\end{eqnarray}
and similarly for $\ps A=i\{P,A\}$.
Any particular bracket equation of motion can then be computed
in either formulation using only
equal-$\s$ brackets,  and these results
agree with (\ref{dotprime}).

\vskip .2cm
\noindent \underline{Coordinate space}

We turn now to the coordinate-space formulation of the theory.
As a first step, we use the phase-space currents
(\ref{classE2}) and eqs.(\ref{xdot},b) in the form
\begin{equation}
p_{i\m}(\hB)={k\over 2}\et_{ab}e_{i\m}{}^{a,-m}e_{j\n}{}^{bn}\pt x^{j\n},
\qquad
\hat\O_{am}{}^\ys=\et_{ab}e_{i\m}{}^{b,-m} \ps x^{i\m}
\end{equation}
to obtain the simple coordinate-space form of the chiral current modes,
\begin{subequations}
\begin{equation}
E_a(m,\t,\s)=k\et_{ab}e_{i\m}{}^{b,-m}(x(\t,\s))\pa x^{i\m}(\t,\s), \qquad
\bE_a(m,\t,\s))=k\et_{ab}\be_{i\m}{}^{b,-m}(x(\t,\s))\bar\pa x^{i\m}(\t,\s)
\label{coordE}
\end{equation}
\begin{equation}
\bpa E_a(m,\t,\s)=\pa \bE_a(m,\t,\s)=0
\end{equation}
\begin{equation}
\pa={1\over2}(\pt+\ps), \qquad
\bpa={1\over2}(\pt-\ps).
\end{equation}
\end{subequations}
This form of the current modes bears a strong resemblance to the usual
coordinate-space currents
$k\et_{ab}e_i{}^b\pa x^i$, $k\et_{ab}\be_i{}^b\bpa x^i$ of the
conventional WZW model on $G$.

Following the usual canonical density formulation,
we also obtain the action of the two-dimensional field theory on $\hat G$,
\\
\begin{subequations}
\vbox to 20pt{
\begin{equation}
S_{\hat F}=\int d\t d\s \sL_{\hat F}, \qquad
\sL_{\hat F}=p_{i\m}\pt x^{i\m}-\H(\l)
\end{equation} }
\vbox to 52pt{
\begin{eqnarray}
\sL_{\hat F}&=& {k\over8\p}\et_{ab}e_{i\m}{}^{a,-m}e_{j\n}{}^{bm}
 (\pt x^{i\m}\pt x^{j\n}-\ps x^{i\m}\ps x^{j\n})
 +{k\over4\p} \hB_{i\m,j\n}\pt x^{i\m} \ps x^{j\n}
 \nonumber \\
 & & +k\l_{i\m} (\ps x^{i\m}-\et^{ab}e_{a,-m}{}^{i\m}\hO_{bm}{}^\ys)
\label{lagrange}
\end{eqnarray} }
\label{action}
\end{subequations}
where $\H(\l)$ and $\sL_{\hat F}$ include
the constraint (\ref{xprime}) with a Lagrange
multiplier $\l_{i\m}(\t,\s)$.
Sections \ref{wzwf} and \ref{Funmeas} give
alternate forms of this action in terms
of the affine group element $\hat g$.

It is straightforward to check that the
Lagrange equations of motion of this system are equivalent to the
Hamiltonian equations (\ref{xdot},c) and the $\ps x$ constraint (\ref{xprime}).
One does not need to include the $\ps p$ constraint (\ref{pprime})
explicitly in the action formulation; it is
implied by the $\pa_\s x$
constraint, the $\pa_\t x$ equation of motion, and the fact
$\pa_\t\pa_\s x=\pa_\s\pa_\t x$.

On the constrained subspace
it is also straightforward to show that $\ps\sL_{\hat F}=0$,
so the Lagrangian is proportional to its density
$L_{\hat F}=\int d\s\sL_{\hat F}=2\p\sL_{\hat F}$,
in parallel with the Hamiltonian.
One can then check backwards that
\begin{equation}
L_{\hat F}=L_M, \qquad S_{\hat F}=S_M
\label{FMeq}
\end{equation}
on the constrained subspace, where $L_M$ is the mechanical Lagrangian
on $\hat G$ in
(\ref{Baction}).

The action (\ref{action}) is another central result of this paper.  We
remark that it bears a strong resemblance to
the sigma model form (\ref{firsteq}) of
the conventional WZW action, except that our action involves
fields on the affine Lie group, and there is an
additional term to enforce the constraint.
In the following section we discuss rewriting this action as a function
of the affine group element $\hg\in \hat G$,
in analogy  to the $g\in G$ formulation (\ref{firstg}) of
the conventional WZW action.

\section{Trace formulation on the affine group}
\label{traces}
\subsection{Rescaled Killing metric and rescaled traces}
\label{trace1}

We begin by investigating the Killing metric on the affine group $\hat G$.
Recall first the definition of
the Killing metric $\et_{ab}$ on Lie $G$,
\begin{equation}
-f_{ac}{}^d f_{bd}{}^c = \Tr(T_a^{adj} T_b^{adj}) = Q_\psi \et_{ab},
\qquad Q_\psi=\psi^2 \tilde{h}
\end{equation}
where $f_{ab}{}^c$, $\tilde h$ and $\psi$ are
respectively the structure constants, the dual Coxeter number and
the highest root of Lie $g$.

In the same way,
the formal Killing metric $\et_{MN}$ on $\hat G$ is defined
by the relation
\begin{equation}
-f_{ML}{}^P f_{NP}{}^L = \Tr(\hT_{M}^{adj}\hT_{N}^{adj})
=\hat Q \et_{MN}
\label{formKill}
\end{equation}
where $f_{LM}{}^N$ are the affine structure constants and
$\hT^{adj}$ is the adjoint matrix representation of the affine algebra in
(\ref{tadj}).  In (\ref{formKill}),
the formal trace $\Tr$
is in fact a sum over the reduced carrier space $am$ because
$(\hT_M^{adj})_\ys{}^N=0$.
By explicit computation, one finds that
\begin{subequations}
\begin{equation}
\Tr(\hT_\ys^{adj} \hT_L^{adj})=\Tr(\hT_L^{adj}\hT_\ys^{adj})=0
\end{equation}
\begin{equation}
\Tr(\hT_{a_1m_1}^{adj}\hT_{a_2m_2}^{adj})
= \hTr(\hT_{a_1m_1}^{adj}\hT_{a_2m_2}^{adj})
\left(\sum_{m\in\Z}\right)
\label{infinite}
\end{equation}
%\vspace{1pt}
%\vspace{-1pt}
\begin{equation}
\hTr(\hT_{am}^{adj}\hT_{bn}^{adj})\equiv\d_{m+n,0}\Tr(T_a^{adj}T_b^{adj})
=Q_\psi \et_{ab}\d_{m+n,0}
\label{reddef}
\end{equation}
\end{subequations}
so the formal traces and the product $\hat Q \et$
have a divergent factor, which is the sum
over modes in (\ref{infinite}).
The rescaled trace $\hTr$ in (\ref{reddef})
and the corresponding rescaled Killing metric $\het_{LM}$
\begin{subequations}
\begin{equation}
\hTr(\hT_L^{adj}\hT_M^{adj})= Q_\psi \het_{LM}
\label{rescaled}
\end{equation}
\begin{equation}
\het_{am,bn}=\et_{ab}\d_{m+n,0}, \qquad \het_{\ys L}=\het_{L\ys}=0
\end{equation}
\label{kmetric}
\end{subequations}
are finite, however, and
the rescaled Killing metric, introduced in Section \ref{ALA},
 has been used many times in the development of the previous
sections.

In fact, the rescaled trace and rescaled Killing metric continue to
be sufficient for
our purposes.  We recall from Section \ref{ALA}
that the rescaled Killing
metric can be used to lower indices and, in particular,
one finds the completely
antisymmetric structure constants,
\begin{equation}
f_{am,bn,cp}=f_{am,bn}{}^{dq}\het_{dq,cp}=f_{abc}\d_{m+n+p,0}
\end{equation}
where the structure constants $f_{am,bn}{}^{dq}$ are given in
(\ref{structure}).
Because it vanishes on the $\ys$ subspace,
the rescaled metric is not invertible on the full space ($am,\ys$).
However, an inverse exists on the $am$ subspace,
\begin{subequations}
\begin{equation}
\het^{am,bn}=\et^{ab}\d_{m+n,0}
\end{equation}
\begin{equation}
H={1\over2k}\het^{am,bn}(E_a(m)E_b(n)+\bar E_a(m)\bar E_b(n))
\label{wzwKill}
\end{equation}
\end{subequations}
and, following its natural occurrence in the WZW
Hamiltonian (\ref{wzwKill}),
this inverse has been used many times in the development of the
previous sections.

The divergent factor in (\ref{infinite}) is not unique to the trace of the
adjoint representation $\hT^{adj}$.  Indeed, the same factor will
recur in the traces over any other matrix representation which
is faithful to the mode structure of the affine algebra.

A large class of such representations can easily be constructed.
For each  matrix irrep $(T_a)_I{}^J, I,J=1\ldots\dim T$ of Lie $g$,
\begin{equation}
[T_a,T_b]=if_{ab}{}^c T_c, \qquad \Tr(T_aT_b)=\chi(T)\et_{ab}
\end{equation}
one has the corresponding matrix representation $\hT(T)$
of affine $g$,
\begin{subequations}
\begin{equation}
(\hT_{am}(T))_{In}{}^{Jp}\equiv\d_{m+n,p}(T_a)_I{}^J, \qquad
(\hT_\ys(T))_{In}{}^{Jp}\equiv 0
\end{equation}
\begin{equation}
[\hT_L(T),\hT_M(T)]=if_{LM}{}^N \hT_N(T)
\end{equation}
\label{trep}
\end{subequations}
where the carrier space of $\hT(T)$ is $(Im)$.
When one takes a naive trace of a product of $N$ of these matrices, one
finds
\begin{subequations}
\begin{equation}
\Tr(\hT_{a_1m_1}(T)\cdots\hT_{a_Nm_N}(T)) =
\hat{\Tr}(\hT_{a_1m_1}(T)\cdots\hT_{a_Nm_N}(T)) \left(\sum_{m\in\Z}\right)
\end{equation}
\begin{eqnarray}
\hat{\Tr}(\hT_{a_1m_1}(T)\cdots\hT_{a_Nm_N}(T)) & \equiv &
\sum_I(\hT_{a_1m_1}(T)\cdots\hT_{a_Nm_N})_{Im}{}^{Im}, \qquad \forall m\in\Z
\nonumber \\
&=& \d_{m_1+\ldots+m_N,0}\Tr(T_{a_1}\cdots T_{a_N})
\label{tracehat}
\end{eqnarray}
\begin{equation}
\hTr(\hT_{am}(T)\hT_{bn}(T))=\chi(T)\het_{am,bn}
\label{treta}
\end{equation}
\end{subequations}
in parallel to eq.(\ref{kmetric}).
Again, the rescaled traces in
(\ref{tracehat}) are finite and sufficient for our purposes.
We will also need the more general relation
\begin{equation}
\hTr(\F(\hT(T))=\sum_I (\F(\hT(T))_{Im}{}^{Im}, \qquad \forall m\in\Z
\end{equation}
which holds for any matrix-valued power series $\F$.

Note that we now have two affine representations, $\hT^{adj}$ and
$\hT(T^{adj})$, corresponding to the adjoint representation of Lie $g$.
These two representations are closely related
\begin{subequations}
\begin{equation}
(\hT_{am}^{adj})_{bn}{}^{cp}=(\hT_{am}(T^{adj}))_{bn}{}^{cp}
= -if_{am,bn}{}^{cp}
\end{equation}
\begin{equation}
(\hT_{am}^{adj})_{bn}{}^\ys=-m\het_{am,bn}
\end{equation}
\end{subequations}
although $\hT^{adj}$ has the extra dimension $\ys$ in its carrier
space.  The traces of the two representations are however the same
\begin{subequations}
\begin{equation}
\hTr(\hT^{adj}_{a_1m_1}\cdots\hT^{adj}_{a_Nm_N})=
\hat{\Tr}(\hT_{a_1m_1}(T^{adj})\cdots\hT_{a_Nm_N}(T^{adj}))
\end{equation}
\begin{equation}
\hTr(\hT_{am}^{adj}\hT_{bn}^{adj})=
\hTr(\hT_{am}(T^{adj})\hT_{bn}(T^{adj}))= \chi(T^{adj})\het_{am,bn}, \quad
\chi(T^{adj})=Q_\psi
\end{equation}
\end{subequations}
because $(T_M^{adj})_\ys{}^N=0$. In what follows, we use the unified
notation $\hT$ to denote any one of the representations
$\hT(T)$ or $\hT^{adj}$.

In matrix representation $\hT$, the group element
$\hga\in\hat G$ equals the reduced group element~$\hg$,
\begin{equation}
\hat\g(\hT,x,y)=\hg(\hT,x)= \exp(\b^{am}(x)\hT_{am})
\end{equation}
because $\hT_\ys=0$ replaces the level in these representations.  Then
one obtains $\hg(\hT)$ analogues of the basic relations which we obtained
for $\hg(J)$ in Section \ref{ALA}.
For example, one has the $\hg(\hT)$ analogue of
eqs.(\ref{unbarred2},b),
\begin{subequations}
\begin{equation}
e_{i\m}(\hT)= -i\hg^{-1}(\hT)\pa_{i\m}\hg(\hT)=e_{i\m}{}^{am}\hT_{am}
\end{equation}
\begin{equation}
\bar e_{i\m}(\hT) =
-i\hg(\hT)\pa_{i\m}\hg^{-1}(\hT)=\be_{i\m}{}^{am}\hT_{am}
\end{equation}
\label{Tviel}
\end{subequations}
where the vielbeins $e_{i\m}{}^{am}(x), \be_{i\m}{}^{am}(x)$, being
representation independent, are the same as those above.

Many, but not all, of these relations can be obtained, as in
(\ref{Tviel}), by the map
$J\rightarrow \hT$ and $k\rightarrow 0$.  An important exception involves the
operator form (\ref{quantE}) of the reduced affine Lie derivatives,
which satisfy
\begin{equation}
E_a(m)\hg(\hT)=\hg(\hT)(\hT_{am}+ke_{am}{}^y), \qquad
\bar E_a(m)\hg(\hT)=-(\hT_{am}-k\bar e_{am}{}^y)\hg(\hT).
\end{equation}
This differs in form from the relation (\ref{indact})
because the reduced affine Lie derivatives are
independent of representation and
explicitly dependent on the level.

\subsection{WZW in terms of $\hg(x(\t,\s))\in\hat G$}
\label{wzwf}
The two-dimensional field theory on $\hat G$ studied in Section \ref{sectt},
\begin{subequations}
\begin{equation}
S_{\hat F}=\int d\t d\s \sL_{\hat F}
\end{equation}
\begin{eqnarray}
\sL_{\hat F}&=& {k\over8\p}\et_{ab}e_{i\m}{}^{a,-m}e_{j\n}{}^{bm}
 (\pt x^{i\m}\pt x^{j\n}-\ps x^{i\m}\ps x^{j\n})
 +{k\over4\p} \hB_{i\m,j\n}\pt x^{i\m} \ps x^{j\n}
 \nonumber \\
 & & +k\l_{i\m} (\ps x^{i\m}-\et^{ab}e_{a,-m}{}^{i\m}\hO_{bm}{}^\ys)
\label{lagrange2}
\end{eqnarray}
\label{action2}
\end{subequations}
is expressed in terms of coordinates on the affine Lie group.
We turn now to rewriting this theory
in terms of rescaled
traces of functions of the reduced affine group element $\hg(\hT,x(\t,\s))$,
using the results of Sections \ref{sectt} and
\ref{trace1}.

Note first that the classical coordinate-space
current modes (\ref{coordE}) can be written in matrix form,
\begin{subequations}
\begin{equation}
E_a(m,\t,\s)\et^{ab}\hT_{b,-m}=-ik\hg^{-1}(\hT,x(\t,\s))\pa\hg(\hT,x(\t,\s))
\end{equation}
\begin{equation}
\bE_a(m,\t,\s)\et^{ab}\hT_{b,-m}=-ik\hg(\hT,x(\t,\s))
\bar\pa\hg^{-1}(\hT,x(\t,\s))
\end{equation}
\label{matmod}
\end{subequations}
using (\ref{Tviel}).
Correspondingly, the current modes can be written as
rescaled traces,
\begin{subequations}
\begin{equation}
E_a(m,\t,\s)=-{i\over\chi(T)}k\hTr\left(\hT_{am}
\hg^{-1}(\hT,x(\t,\s))\pa\hg(\hT,x(\t,\s))\right)
\end{equation}
\begin{equation}
\bE_a(m,\t,\s)=-{i\over\chi(T)}k
\hTr\left(\hT_{am}\hg(\hT,x(\t,\s))\bar\pa\hg^{-1}(\hT,x(\t,\s))\right)
\end{equation}
\end{subequations}
in any matrix representation $\hT$, using (\ref{treta}).
For brevity below, we write $\hg$ for $\hg(\hT,x(\t,\s))$ and
$\chi$ for $\chi(T)$.

Following the example of the currents, the kinetic terms in the
action (\ref{action2}) can be written
\begin{equation}
{k\over8\p}\et_{ab}e_{i\m}{}^{a,-m}e_{j\n}{}^{bm}
(\pt x^{i\m}\pt x^{j\n} - \ps x^{i\m}\ps x^{j\n}) =
-{k\over2\p\chi}
\hat{\Tr}\left(\hg^{-1}\pa\hg
\hg^{-1}\bar\pa\hg\right)
\label{kinetic}
\end{equation}
using eq.(\ref{Tviel}).

We next consider the $\hB$ term in the action.  Note
first that the antisymmetric tensor field
$\hB$, defined in (\ref{Bdef}), can be written as a
rescaled trace
{\samepage
\begin{subequations}
\begin{equation}
\hB_{i\m,j\n}(x(\t,\s))={i\over\chi}\int_0^1 dt\, \hTr \left(
\hH\pa_{[i\m}e^{-it\hH}\pa_{j\n]}e^{it\hH} \right), \qquad
\hH=\b^{am}(x(\t,\s))\hT_{am}
\label{Bint}
\end{equation}
\begin{equation}
X_{[i\m}Y_{j\n]}\equiv X_{i\m}Y_{j\n}-X_{j\n}Y_{i\m}
\end{equation}
\end{subequations}
}
in any $\b$-basis. The equality of (\ref{Bdef})
and (\ref{Bint}) follows by integration over the parameter $t$.
Using the integral representation (\ref{Bint}), one also verifies the
cyclic identity
\begin{equation}
\pa_{i\m}\hB_{j\n,k\rho} +
\pa_{j\n}\hB_{k\rho,i\m} + \pa_{k\rho}\hB_{i\m,j\n} =
{i\over\chi}\hTr(e_{i\m}[e_{j\n},e_{k\rho}]).
\label{cyclic}
\end{equation}
Both identities (\ref{Bint}) and (\ref{cyclic}) are analogous to
standard relations (see Appendix A) satisfied by the conventional antisymmetric
tensor field $B_{ij}$ on Lie $g$.

The rest of the discussion on $\hat G$ parallels the usual development of the
WZW term on $G$.
The cyclic identity can be used to integrate the following three-form on
$\hat G$,
\begin{subequations}
\begin{equation}
\hTr(\hg^{-1}d\hg)^3 \equiv
d^3\xi\,
\vare^{ABC}\hTr(\hg^{-1}\pa_A\hg\hg^{-1}\pa_B\hg\hg^{-1}\pa_C\hg)
=d^3\xi\,\pa_A \W^A
\end{equation}
\begin{equation}
\W^A\equiv-{3\over2}\chi \vare^{ABC} \pa_B x^{i\m} \pa_C x^{j\n}\hB_{i\m,j\n}
\end{equation}
\begin{equation}
A=(\t,\s,\rho), \qquad
d^3\xi=d\t d\s d\rho, \qquad
\vare^{012}=+1
\end{equation}
\begin{equation}
\int\hTr(\hg^{-1}d\hg)^3=\int d\t d\s \W^\rho(\t,\s,\rho=1)
=-3\chi \int d\t d\s \pt x^{i\m}\ps x^{j\n} \hB_{i\m,j\n}
\end{equation}
\label{diverg}
\end{subequations}
where $0\le\rho\le1$ is the radial coordinate of the usual WZW cylinder,
and the two-dimensional $\hat g$ in (\ref{matmod}-19) is the boundary
value of this $\hg$ at radius $\rho=1$.

Using (\ref{diverg}), we obtain our first $\hg$ form of the two-dimensional
field theory on $\hat G$,
\begin{eqnarray}
S_{\hat F}&=&-{k\over2\p\chi}\int d\t d\s
\hat{\Tr}\left(\hg^{-1}\pa\hg
\hg^{-1}\bar\pa\hg
 \right)
 - {k\over 12\p \chi} \int \hTr(\hg^{-1}d\hg)^3
\nonumber \\
& &
 +{k\over\chi}\int d\t d\s
\hTr( \hT_{am}\ln \hg)
 (i\ps+m)\l^{am}
\label{traction}
\end{eqnarray}
where the second term is a three-dimensional WZW term on the affine group.
The last term in (\ref{traction}) is the constraint term, whose form
follows from eq.(\ref{bconst}).
The multiplier $\l^{am}$ is
defined by the invertible relation
$\l_{i\m}=\pa_{i\m}\b^{am}\het_{am,bn}\l^{bn}$
and a sum on $m\in\Z$ is understood in this term.

The first two terms of the
action (\ref{traction}), without the constraint term, were considered
as a theory
in Ref.~\cite{schwim}, and the theory was found to have an infinite degeneracy
in that case.
In the formulation (\ref{traction}),
it is the role of the constraint to remove that
degeneracy and to implement the
classical equivalence with the conventional formulation
of the WZW model on the Lie
group.
In Section \ref{Funmeas}, we find a more elegant form of the constraint
term which is
associated to the formal quantum equivalence of this formulation with
the conventional formulation.

\newpage
\part{Translation dictionary: $\hat G \leftrightarrow G$}
\label{part3}

\section{Strategy: Partial transmutation of target and base spaces}

In Part \ref{part2}, we found two actions,
(\ref{myaction},8) and (\ref{traction}), on the affine
Lie group $\hat G$ which must be equivalent to the conventional
formulation of the WZW model on
the Lie group $G$.
This equivalence is clear because the Hamiltonians of the theories are
isomorphic.

In this part, we find the explicit translation dictionary between the
formulations on $\hat G$ and $G$, which shows that our new actions on $\hat G$
are mode formulations of WZW.

The overall picture which emerges is a quartet of equivalent formulations
of WZW theory,
\begin{equation}
\begin{picture}(200,90)(0,80)
\put(5,150){$\hat g(x(\tau))$}
\put (25,140){\vector(0,-1){50}}
\put (25,135){\vector(0,1){5}}
\put(0,80){$g(x(\tau,\sigma))$}
\put(45,153){\vector(1,0){80}}
\put(50,153){\vector(-1,0){5}}
\put(130,150){$\hat g(x(\tau,\sigma))$}
\put(60,160){constraint}
\put(150,140){\vector(0,-1){50}}
\put(150,135){\vector(0,1){5}}
\put(125,80){$g(x(\tau,\sigma,\ts))$}
\put(-12,115){modes}
\put(113,115){modes}
\put(52,83){\vector(1,0){69}}
\put(62,83){\vector(-1,0){10}}
\put(60,90){constraint}
\end{picture}
\label{picture}
\end{equation}
where $\hg\in\hat G$ and $g\in G$.
The conventional formulation of WZW on $G$ is in the
lower left.
The top line of the picture describes the two formulations
on $\hat G$ in Part \ref{part2}, whose equivalence
to each other was discussed in Section \ref{sectt}.
More generally, the horizontal direction of the picture shows the use
of constraints to change the dimension of the base space (see also Section
\ref{wzw3}).
The left column describes the equivalence (which is studied in
Sections~\ref{modestolocal}-\ref{WZWequiv}) of the
mechanical formulation of WZW on $\hat G$ with the conventional
formulation on $G$.

The central relation underlying the equivalence in the
left column is the
mode identity
\begin{equation}
\b^a(x^i(\t,\s))=
\sum_m e^{im\s}\b^{am}(x^{i\m}(\t))
\label{modident}
\end{equation}
where $x^i(\t,\s)$ is the local WZW coordinate on $G$, and
the tangent-space coordinates
$\b^{a}$ and $\b^{am}$ appear in the group elements of $G$ and
$\hat G$ as
\begin{equation}
g(T,x(\t,\s))=e^{i\b^a(x(\t,\s))T_a}, \qquad
\hg(\hT,x(\t))=e^{i\b^{am}(x(\t))\hT_{am}}\;.
\label{gbet}
\end{equation}
As in Section \ref{traces},
$\hT$ is any of the affine matrix representations
$\hT(T)$ or $\hT^{adj}$,
and $T$ is the corresponding
matrix representation of Lie $g$.

The mode identity (\ref{modident}) emphasizes the fact
that the operation of moding is a partial transmutation
\begin{subequations}
\begin{equation}
x^{i\m}(\t)\leftrightarrow x^i(\t,\s)\qquad
\end{equation}
\begin{equation}
\hg(x(\t)):\;\; \R\mapsto\hat G, \qquad\;\;
g(x(\t,\s)):\;\; \R\times S^1 \mapsto G
\end{equation}
\end{subequations}
between the target space and the base space.
More generally, such partial transmutations operate
in the vertical direction of the
picture, and the transmutation
in the right column allows us
to express the field-theoretic formulation on $\hat G$ as a
three-dimensional field theory on $G$ (see Section \ref{wzw3}).

\section{Modes on $\hat G$ and local fields on $G$}
\label{modestolocal}

In this section, we develop the translation dictionary which underlies
the equivalence
\begin{equation}
\begin{picture}(200,90)(0,80)
\put(5,150){$\hat g(x(\tau))$}
\put (25,140){\vector(0,-1){50}}
\put (25,135){\vector(0,1){5}}
\put(0,80){$g(x(\tau,\sigma))$}
\put(130,150){$\hat g(x(\tau,\sigma))$}
\put(125,80){$g(x(\tau,\sigma,\ts))$}
\put(-12,115){modes}
\multiput(65,80)(20,0){3}{$\cdot$}
\multiput(150,100)(0,15){3}{$\cdot$}
\multiput(65,150)(20,0){3}{$\cdot$}
\end{picture}
\end{equation}
described by the left column in the picture (\ref{picture}).
In Section \ref{wzw3}, the
dictionary will also be applied to the right column of the picture.

As a conceptual orientation, we recall first that
the operator current modes (\ref{quantE2}),
\begin{subequations}
\begin{equation}
E_a(m)=-ie_{am}{}^{i\m}\cD_{i\m}(\hB)
+{1\over2}k\hO_{am}{}^\ys, \qquad
\bE_a(m)=-i\be_{am}{}^{i\m}\cD_{i\m}(\hB)
-{1\over2}k\hO_{am}{}^\ys
\end{equation}.
\begin{equation}
\cD_{i\m}(\hB)=
\pa_{i\m}-{i\over2}k\hB_{i\m,j\n}e_{a,-m}{}^{j\n}\et^{ab}\hO_{bm}{}^\ys
\end{equation}
\label{quantagain}
\end{subequations}
and the corresponding mechanical Lagrangian (\ref{Baction})
on $\hat G$,
\begin{equation}
L_M={k\over4}\et_{ab}e_{i\m}{}^{a,-m}e_{j\n}{}^{bm} \pt x^{i\m}\pt x^{j\n}
  - {k\over4} \left( \hO_{am}{}^\ys+2\hB_{i\m,j\n}
    \pt x^{j\n}e_{am}{}^{i\m}\right)\et^{ab}\hO_{b,-m}{}^\ys
\label{Bform2}
\end{equation}
make no reference to any spatial variable $\s$.
In what follows, we will define $\s$-dependent local fields on $G$
whose $\s$-independent modes are the quantities on $\hat G$.
Except where it is relevant, we will suppress the time dependence of
all quantities.
We will discuss the translation dictionary first for the quantum
system, returning later to indicate the simple changes necessary
for the corresponding classical results.

We begin with the canonical
operator system
\begin{equation}
[x^{i\m},p_{j\n}]=i\d_{j\n}{}^{i\m}, \qquad
p_{i\m}=-i\pa_{i\m}
\end{equation}
where $x^{i\m}$ are the Einstein coordinates on $\hat G$.  It is useful
to define the corresponding canonical tangent-space system,
\begin{subequations}
\begin{equation}
p_{am} \equiv -i {\pa\over\pa\b^{am} }
\equiv -i\pa_{am} = (\pa_{am}x^{i\m}) p_{i\m}
\label{betmom}
\end{equation}
\begin{equation}
[\b^{am},p_{bn}]=i \d_{bn}{}^{am}, \qquad
[\b^{am},\b^{bn}]=[p_{am},p_{bn}]=0
\label{pbb}
\end{equation}
\label{betsys}
\end{subequations}
where $\b^{am}$ is the tangent-space coordinate on $\hat G$ in (\ref{gbet}) and
$\pa_{am}x^{i\m}(\b)$ is the inverse of the
matrix $\pa_{i\m}\b^{am}(x)$.

We then define the local WZW fields in terms of a periodic
coordinate $0\le\s<2\p$,
\begin{subequations}
\begin{equation}
\b^a(x(\s))\equiv\sum_m e^{im\s} \b^{am}(x)
\label{Fourb}
\end{equation}
\begin{equation}
p_a(\s)\equiv{1 \over 2\p}\sum_m e^{-im\s} p_{am}
\label{Fourp}
\end{equation}
\label{Fourbp}
\end{subequations}
where $x^i(\s), i=1\ldots\dim g$
are the local Einstein coordinates on $G$
and $\b^a(x(\s))$ are the tangent-space coordinates on $G$ in (\ref{gbet}).
Note that in this moding upper and lower tangent-space indices are
associated to $e^{im\s}$ and $e^{-im\s}$ respectively.

As we will see below, the tangent-space mode identities
(\ref{Fourbp}) are correct because: \\
\bul The mode identities guarantee the same simple
$e^{\pm im\s}$ moding for all
objects with tangent-space
or carrier-space indices.  The moding of objects with Einstein
indices is more involved, as discussed
below. \\
\bul The mode identities guarantee equivalence of the mechanical system
(\ref{Bform2}) on $\hat G$ with the conventional WZW model on $G$.
We mention in particular (see Section \ref{WZWequiv}) that
conventional WZW averages on $G$
\begin{equation}
\langle \F[\b^a(x(\t,\s))]\rangle\raisebox{-2pt}{${}_G$} =
\langle \F[\sum_m e^{im\s}\b^{am}(x(\t))]\rangle_{\hat G_M}
\label{ggtrans}
\end{equation}
can be computed for any $\F$ as shown, using the mechanical
formulation ($\hat G_M$) on $\hat G$.

The relation (\ref{Fourb}) allows independent choices of bases on $\hat G$
and on $G$, which control the moding of the Einstein coordinates.
For example, one may choose the standard bases
\begin{equation}
\b^a(x(\s))=x^i(\s)e_i{}^a(0), \qquad
\b^{am}(x)=x^{i\m}e_{i\m}{}^{am}(0)
\end{equation}
where $e_i{}^a(0)$ and $e_{i\m}{}^{am}(0)$ are
the vielbeins at the origin on $G$ and $\hat G$ respectively.
Then one obtains the mode relation of the Einstein coordinates,
\begin{equation}
x^i(\s)= x^{j\n}(\sum_m e_{j\n}{}^{am}(0)
e^{im\s})e_a{}^i(0)
\label{badmod}
\end{equation}
and more complicated relations are generally obtained for other
basis choices.

It is straightforward to check that the local tangent-space
fields in (\ref{Fourbp})
 are canonical
\begin{equation}
[\b^{a}(x(\s)),p_b(\s')]=i \d_b{}^a\d(\s-\s'), \qquad
[\b^a(x(\s)),\b^b(x(\s'))]=[p_a(\s),p_b(\s')]=0
\label{lbrack}
\end{equation}
so $p_a(\s)$ is the functional derivative $-i\d/\d\b^a(x(\s))$.

Although their moding may be complicated, the Einstein coordinates
are periodic functions of $\s$, and one can
also find the corresponding local
canonical Einstein system,
\begin{subequations}
\begin{equation}
p_i(\s)\equiv \pa_i\b^a(x(\s)) p_a(\s)
\end{equation}
\begin{equation}
[x^i(\s),p_j(\s')]=i\d_j{}^i \d(\s-\s'), \qquad
[x^i(\s),x^j(\s')]=[p_i(\s),p_j(\s')]=0
\end{equation}
\end{subequations}
by using chain rule from the local tangent-space system (\ref{lbrack}).
It follows that the local Einstein momenta are functional derivatives
\begin{eqnarray}
p_i(\s)&=&{1\over2\p}
\pa_i\b^a(x(\s))\sum_m e^{-im\s}(\pa_{am}x^{i\m}(\b))i\pa_{i\m}
\nonumber \\
&=& -i {\d\over\d x^i(\s)}
\label{pfunct}
\end{eqnarray}
with respect to the local Einstein coordinates.

\vskip .2cm
\noindent
\underline{Group elements}

We consider next the the group elements
$\hg\in\hat G$ and $g\in G$ in (\ref{gbet}),
whose mode relations have the form
\begin{subequations}
\begin{equation}
\sum_m e^{i(n-m)\s}\hg(\hT,x)_{Im}{}^{Jn}= g(T,x(\s))_I{}^J, \qquad
\forall n\in\Z
\label{uhhuh}
\end{equation}
\begin{equation}
\hg(\hT,x)_{Im}{}^{Jn}={1\over2\p}\int d\s e^{i(m-n)\s}g(T,x(\s))_I{}^J
\label{invg}
\end{equation}
\label{gmod}
\end{subequations}
where $I,J=1\ldots\dim T$.
The inverse relation (\ref{invg}) follows from (\ref{uhhuh}).

The relation (\ref{uhhuh}) is proven separately for each
order in $\hH$ and $H$, where
{\samepage
\begin{subequations}
\begin{equation}
\hH\equiv\b^{am}(x)\hT_{am}, \qquad
\hg(\hT,x)=e^{i\hH}
\end{equation}
\begin{equation}
H(\s)\equiv\b^a(x(\s))T_a, \qquad
g(T,x(\s))=e^{iH(\s)}.
\end{equation}
\end{subequations}
}
As an illustration, we discuss the lowest orders explicitly.
To zeroth order, the relation (\ref{uhhuh}) is an identity because
\begin{equation}
(\one)_{Im}{}^{Jn}=\d_I{}^J\d_m{}^n, \qquad
(\one)_I{}^J=\d_I{}^J.
\end{equation}
The first order computation is
\begin{eqnarray}
\sum_m e^{i(n-m)\s}\hH_{Im}{}^{Jn} &=&
\sum_{p,m}e^{i(n-m)\s}\b^{ap}(x)(\hT_{ap})_{Im}{}^{Jn} \nonumber\\
&=& \sum_{p,m}e^{i(n-m)\s}\b^{ap}(x)(T_a)_I{}^J\d_{p+m,n} \nonumber\\
&=& \sum_p e^{ip\s}\b^{ap}(x)(T_a)_I{}^J \nonumber\\
&=& \b^a(x(\s))(T_a)_I{}^J \nonumber\\
&=& H(\s)_I{}^J
\end{eqnarray}
and higher orders
are easily checked following similar steps.

The form of the result (\ref{uhhuh}) is somewhat surprising,
because one might have expected a double sum over $m$ and $n$.
In fact this form is natural because, in their carrier space indices,
the affine matrix representations
$\hT_{Im}{}^{Jn}$ (see (\ref{trep})) and the affine group elements
$\hg_{Im}{}^{Jn}$ are functions only of $m-n$.

In the same way, one establishes the general relations
\begin{subequations}
\begin{equation}
\sum_m e^{i(n-m)\s} (\F(\hH))_{Im}{}^{Jn}=(\F(H(\s)))_I{}^J, \qquad
\forall n\in\Z
\label{twoindex}
\end{equation}
\begin{equation}
\F(\hH))_{Im}{}^{Jn}={1\over2\p}\int d\s e^{i(m-n)\s}(\F(H(\s)))_I{}^J
\end{equation}
\begin{equation}
\hTr(\F(\hH))={1\over2\p}\int d\s \Tr(\F(H(\s)))
\label{trmod}
\end{equation}
\end{subequations}
for all power series $\F(H)$, where $\Tr$ is trace on Lie $g$ and the
rescaled traces $\hTr$ on $\hat G$ are defined in Section \ref{traces}.

\vskip .2cm
\noindent
\underline{Antisymmetric tensor fields}

The general
 relation (\ref{twoindex}) is sufficient to obtain the mode relations
of any
object with two free indices, such as the group elements in (\ref{gmod}).
As a second example, we consider the mode relation between the
antisymmetric tensor field $\hB_{i\m,j\n}$ on $\hat G$ and the
standard antisymmetric tensor field $B_{ij}$ on $G$.  Since these
objects have Einstein indices, we employ the transition functions
$\pa_{am}x^{i\m}(\b)$ and $\pa_a x^i(\b(\s))$ to generate tangent-space
structures with simple moding,
\begin{subequations}
\begin{equation}
\pa_{am}x^{i\m}(\b)\hB_{i\m,j\n}(x)\pa_{bn}x^{j\n}(\b)=
(N(\hHa))_{am}{}^{cp} \het_{cp,bn}
\end{equation}
\begin{equation}
\pa_a x^i(\b(\s))B_{ij}(x(\s))\pa_b x^j(\b(\s))=
(N(\Ha(\s)))_a{}^c \et_{cb}
\end{equation}
\begin{equation}
\sum_m e^{i(n-m)\s} (N(\hHa))_{am}{}^{bn} = (N(\Ha(\s)))_a{}^b, \qquad
\forall n\in\Z
\label{Nmod}
\end{equation}
\begin{equation}
\hHa\equiv\b^{am}(x)\hT^{adj}_{am}, \qquad
\Ha(\s)\equiv\b^a(x(\s))T^{adj}_a
\end{equation}
\end{subequations}
where the function $N(H)$, which is the same for both $\hat G$ and $G$, is
given explicitly in Appendices A and B\@.  The mode relation
(\ref{Nmod}) is a special case of (\ref{twoindex}), and, using this
relation, we find that
\begin{equation}
\sum_m e^{-i(m+n)\s}
\pa_{am}x^{i\m}(\b)\hB_{i\m,j\n}(x)\pa_{bn}x^{j\n}(\b)=
\pa_a x^i(\b(\s))B_{ij}(x(\s))\pa_b x^j(\b(\s)).
\label{hello1}
\end{equation}
This mode relation may also be written
\begin{equation}
B_{ij}(x(\s))=\pa_i\b^a(x(\s) )
\left(\sum_m e^{-i(m+n)\s} \pa_{am}x^{k\m}(\b)\hB_{k\m,l\n}(x)
 \pa_{bn}x^{l\n}(\b)\right) \pa_j\b^b(x(\s))
\label{hello2}
\end{equation}
because the transition functions are invertible.  In both
eqs.(\ref{hello1}) and (\ref{hello2}), $n\in\Z$ is arbitrary.

\vskip .2cm \noindent
\underline{Vielbeins}

As another example with two indices, we follow similar steps to obtain
the mode
relations of the vielbeins on $\hat G$ and $G$,
\begin{subequations}
\begin{eqnarray}
\sum_m e^{i(n-m)\s} \pa_{am}x^{i\m}(\b)e_{i\m}{}^{bn}(x) &=&
\sum_m e^{i(n-m)\s} (M(\hHa))_{am}{}^{bn} \\
&=& (M(\Ha(\s)))_a{}^b \\
&=& \pa_a x^i(\b(\s))e_i{}^b(x(\s))
\end{eqnarray}
\begin{equation}
e_i{}^b(x(\s))= \pa_i\b^a(x(\s))
\sum_m e^{i(n-m)\s} \pa_{am}x^{j\n}(\b)e_{j\n}{}^{bn}(x), \qquad
\forall n\in\Z
\end{equation}
\label{vielmod}
\end{subequations}
where the function $M(H)$ is given explicitly
in Appendices A and B\@.
We also find the corresponding  results
\begin{subequations}
\begin{eqnarray}
e_b{}^i(x(\s))&=&\sum_m e^{-i(n-m)\s}e_{bn}{}^{j\n}(x)\pa_{j\n}\b^{am}(x)
\pa_a x^i(\b(\s))  \\
\be_i{}^b(x(\s))&=& \pa_i\b^a(x(\s))
\sum_m e^{i(n-m)\s} \pa_{am}x^{j\n}(\b)\be_{j\n}{}^{bn}(x)  \\
\be_b{}^i(x(\s))&=&\sum_m e^{-i(n-m)\s}\be_{bn}{}^{j\n}(x)\pa_{j\n}\b^{am}(x)
\pa_a x^i(\b(\s)),
\qquad \forall n\in\Z
\end{eqnarray}
\end{subequations}
for the other vielbein and the inverse
vielbeins.

\vskip .2cm \noindent
\underline{Other mode relations}

We turn now to mode relations for general objects
with fewer than two free indices.
These relations
 will be important for the currents and the action, discussed in Sections
\ref{lflc} and \ref{WZWequiv}.
In what follows,
we limit ourselves to the adjoint representations.

A general class of one-index tangent-space objects is formed by contraction of
the tangent-space coordinates or tangent-space momenta with the general
two-index structure $\F(H)$ in (\ref{twoindex}).  For example, one has
\begin{eqnarray}
\b^a(x(\s)) (\F(\Ha(\s)))_a{}^b  &=&
\sum_m e^{im\s} \b^{am}(x) \sum_n e^{i(n-m)\s} (\F(\hHa))_{am}{}^{bn}
\nonumber \\
&=& \sum_n e^{in\s} \b^{am}(x) (\F(\hHa))_{am}{}^{bn} .
\label{oneindex}
\end{eqnarray}
Similarly, one finds the relations
\begin{subequations}
\begin{equation}
\sum_n e^{in\s} \pt\b^{am}(x) (\F(\hHa))_{am}{}^{bn} =
\pt\b^a(x(\s)) (\F(\Ha(\s)))_a{}^b
\end{equation}
\begin{equation}
\sum_n e^{in\s}im\b^{am}(x) (\F(\hHa))_{am}{}^{bn}=
\pa_\s\b^a(x(\s)) (\F(\Ha(\s)))_a{}^b
\label{bsrel}
\end{equation}
\begin{equation}
\sum_m e^{-im\s} (\F(\hHa))_{am}{}^{bn}p_{bn}=
2\p (\F(\Ha(\s)))_a{}^b p_b(\s)
\label{prel}
\end{equation}
\end{subequations}
by using $\pt\b$, $\ps\b$ or $p$ instead of $\b$ in (\ref{oneindex}).

As an application of (\ref{bsrel}), we find the relations
\begin{subequations}
\begin{equation}
\sum_m e^{-im\s}\hg(\hT^{adj},x)_{am}{}^\ys =
\ps x^i(\s)\be_i{}^b(x(\s))\et_{ba}
\end{equation}
\begin{equation}
\sum_m e^{-im\s}\hO_{am}{}^\ys(x)=\ps x^i(\s)e_i{}^b(x(\s))\et_{ba}
\label{ommod}
\end{equation}
\end{subequations}
where we have also used the explicit $\b$-basis forms of the
adjoint action $\hO=\hg^{-1}(\hT^{adj})$ and the vielbeins given in
the appendices.
Taken with eq.(\ref{gmod}), these results complete the mode relations of the
affine group elements.

A large class of mode
relations for objects with no free indices is similarly constructed.
We give here only some representative results,
\begin{subequations}
\begin{equation}
\b^{am}(x)(\F(\hHa))_{am}{}^{cp}\het_{cp,bn}\b^{bn}(x) =
{1\over2\p}
\int d\s \b^a(x(\s))(\F(\Ha(\s)))_a{}^c\et_{cb}\b^{b}(x(\s))
\end{equation}
\begin{equation}
\pt\b^{am}(x)(\F(\hHa))_{am}{}^{cp}\het_{cp,bn}\pt\b^{bn}(x)
={1\over2\p}
\int d\s \pt\b^a(x(\s))(\F(\Ha(\s)))_a{}^c\et_{cb}\pt\b^{b}(x(\s))
\end{equation}
\begin{equation}
\pt\b^{am}(x)(\F(\hHa))_{am}{}^{cp}\het_{cp,bn}in\b^{bn}(x)
={1\over2\p}
\int d\s \pt\b^a(x(\s))(\F(\Ha(\s)))_a{}^c\et_{cb}\ps\b^{b}(x(\s))
\end{equation}
\begin{equation}
m\b^{am}(x)(\F(\hHa))_{am}{}^{cp}\het_{cp,bn}n\b^{bn}(x)
=-{1\over2\p}
\int d\s \ps\b^a(x(\s))(\F(\Ha(\s)))_a{}^c\et_{cb}\ps\b^{b}(x(\s))
\end{equation}
$$
im\b^{am}(x)(\F_1(\hHa))_{am}{}^{dq} \b^{bn}(x)(\F_2(\hHa))_{bn}{}^{er}
\pt\b^{cp}(x)(\F_3(\hHa))_{cp}{}^{fs} f_{dq,er,fs}=
\vphantom{{1\over2\p}}
$$
\begin{equation}
{1\over2\p}\int d\s \ps\b^a(x(\s))(\F_1(\Ha(\s))_a{}^d
\b^b(x(\s))(\F_2(\Ha(\s))_b{}^e \pt\b^c(x(\s))(\F_3(\Ha(\s)))_c{}^f
f_{def}
\label{FFF}
\end{equation}
\label{relations}
\end{subequations}
although other relations of this type are easily obtained to describe
contraction of $\F(H)$ with tangent-space momenta.

As anticipated, the translation dictionary
of this section conforms to
the rule that
 upper and lower (affine) tangent space and carrier space
indices are associated to
$e^{im\s}$ and $e^{-im\s}$ respectively.
This rule holds as well when
(affine) tangent-space indices are raised and lowered with the
rescaled Killing metric.

The translation dictionary has an isomorphic classical
form which is obtained by replacing the
quantum operators $p_{i\m}$, $p_{am}$, $p_i(\s)$, and $p_a(\s)$ with
their corresponding classical momenta.

In the following sections, we apply the translation dictionary to the
central structures of the theory, that is the currents and the action.

\section{Local fields and local currents}
\label{lflc}

The operator current
modes (\ref{quantagain})
\begin{subequations}
\begin{equation}
E_a(m)=-ie_{am}{}^{i\m}\cD_{i\m}(\hB)
+{1\over2}k\hO_{am}{}^\ys, \qquad
\bE_a(m)=-i\be_{am}{}^{i\m}\cD_{i\m}(\hB)
-{1\over2}k\hO_{am}{}^\ys
\end{equation}.
\begin{equation}
\cD_{i\m}(\hB)\equiv
\pa_{i\m}-{i\over2}k\hB_{i\m,j\n}e_{a,-m}{}^{j\n}\et^{ab}\hO_{bm}{}^\ys
\end{equation}
\label{stillmore}
\end{subequations}
are functions of the coordinates on the affine Lie group, and derivatives
with respect to these coordinates.  In this section, we use the
translation dictionary above to express the local operator
currents,
\begin{equation}
E_a(\s)= \sum_m e^{-im\s} E_a(m), \qquad
\bar E_a(\s)= \sum_m e^{-im\s} \bar E_a(m)
\label{yawn}
\end{equation}
as functions of the local fields on the Lie group, and functional
derivatives with respect to these fields.
This section is strictly current-algebraic and does not depend
on any particular
dynamics\footnote{In the context of the mechanical model on $\hat G$,
the classical analogues of the current modes (\ref{stillmore}) were
denoted by $E_a(m,\t)$ and $\bE_a(m,\t)$ in
eq.(\ref{timeEb}). In the context of the field theory on $\hat G$,
the classical analogues of the local currents
(\ref{yawn}) were denoted by $E_a(\t,\s)$ and
$\bE_a(\t,\s)$ in eq.(\ref{locE}).}.

For the left-invariant currents, follow the steps
\begin{subequations}
\begin{eqnarray}
E_a(\s)&=&\sum_m e^{-im\s}
(e_{am}{}^{i\m}(x)p_{i\m}(\hB)
+{1\over2}k\hO_{am}{}^\ys(x)) \\
&=& \sum_m e^{-im\s} (
(C(\hHa))_{am}{}^{bn} p_{bn}
-i k n\b^{bn}(x) (D(\hHa))_{bn}{}^{cp} \het_{cp,am} \label{Eb} )\\
&=&2\p (C(H^{adj}(\s)))_a{}^b p_b(\s)
-k\pa_\s \b^c(x(\s)) (D(H^{adj}(\s)))_c{}^b \eta_{ba} \label{Ec} \\
&=&2\p  e_a{}^i(x(\s))p_i(B,\s)+{k\over2}\et_{ab}e_i{}^b(x(\s))\ps x^i(\s)
\label{Ed}
\end{eqnarray}
\begin{equation}
p_{i\m}(\hB)\equiv -i\cD_{i\m}(\hB)=
p_{i\m}-{k\over2}\hB_{i\m,j\n}(x)
e_{a,-m}{}^{j\n}(x)\et^{ab}\hO_{bm}{}^\ys(x)
\end{equation}
\begin{equation}
p_i(B,\s)\equiv
p_i(\s)-{k\over 4\p}B_{ij}(x(\s))\ps x^j(\s)
\end{equation}
\end{subequations}
where the operator momenta $p_{i\m}$, $p_{am}$, $p_i(\s)$, and $p_a(\s)$
are defined in Section \ref{modestolocal}.
To obtain (\ref{Eb}),
we used the explicit $\b$-basis form of the reduced affine
Lie derivatives given in Appendix B\@.  The
functions $C(H)$ and $D(H)$ are also given in this Appendix.
Eq.(\ref{Ec}) follows from the mode relations (\ref{bsrel}) and
(\ref{prel}).
Finally to obtain (\ref{Ed}), we reorganized (\ref{Ec}) using the explicit
$\b$-basis forms (see Appendix A) of
the vielbein $e_i{}^a$ and the antisymmetric tensor
field $B_{ij}$ on $G$.

Following similar steps for $\bar E$, we summarize the results for
the local operator currents of affine $g\times g$,
\begin{subequations}
\begin{equation}
E_a(\s)=-2\p i e_a(x(\s)){}^i\cD_i(B,\s)+{k\over2}\et_{ab}e_i{}^b(x(\s))
\ps x^i(\s)
\end{equation}
\begin{equation}
\bar E_a(\s)=-2\p i \be_a{}^i(x(\s))
\cD_i(B,\s)-{k\over2}\et_{ab}\be_i{}^b(x(\s))\ps x^i(\s)
\end{equation}
\begin{equation}
\cD_i(B,\s)\equiv ip_i(B,\s)=
{\d\over\d x^i(\s)}-{i\over4\p}kB_{ij}(x(\s))\ps x^j(\s)
\end{equation}
\end{subequations}
where we have used the fact that the local Einstein momentum
$p_i(\s)$ in (\ref{pfunct}) is a
functional derivative.

The classical version of this result is Bowcock's canonical
representation \cite{bow} of affine $g\times g$,
\begin{subequations}
\begin{equation}
E_a(\s)=2\p e_a{}^i(x(\s))p_i(B,\s)+{k\over2}\et_{ab}e_i{}^b(x(\s))
\ps x^i(\s)
\end{equation}
\begin{equation}
\bar E_a(\s)=2\p \be_a{}^i(x(\s))p_i(B,\s)-{k\over2}\et_{ab}\be_i{}^b(x(\s))
\ps x^i(\s)
\end{equation}
\begin{equation}
p_i(B,\s)= p_i(\s)-{k\over4\p}B_{ij}(x(\s))\ps x^j(\s)
\end{equation}
\end{subequations}
where $p_i(\s)$ are classical canonical momenta.

\section{Mechanics on $\hat G$ and WZW on $G$}
\label{WZWequiv}
\subsection{Classical equivalence}
\label{wzwequiv}

In this section, we use the translation dictionary of Section
\ref{modestolocal} to show that the mechanical Lagrangian $L_M$ on $\hat G$
in (\ref{Bform2})
is equal to the conventional WZW Lagrangian,
\begin{subequations}
\begin{eqnarray}
L_M&=&{k\over4}\et_{ab}e_{i\m}{}^{a,-m}e_{j\n}{}^{bm} \pt x^{i\m}\pt x^{j\n}
  - {k\over4} \left( \hO_{am}{}^\ys+2\hB_{i\m,j\n}
    \pt x^{j\n}e_{am}{}^{i\m}\right)\et^{ab}\hO_{b,-m}{}^\ys \quad
\label{Bform3}
\\
&=&{k\over8\p}\int d\s \left[
\et_{ab}e_i{}^a e_j{}^b (\pt x^i\pt x^j - \ps x^i \ps x^j)
 + 2B_{ij}\pt x^i \ps x^j \right]  \equiv L_{WZW}
\label{usualsig}
\end{eqnarray}
\label{neverends}
\end{subequations}
where $L_{WZW}$ in (\ref{usualsig}) is the usual sigma model form of
WZW on $G$.

To see this equality, follow the steps,
\begin{subequations}
\begin{eqnarray}
L_M&=&{k\over 4}(\pt\b^{am}\pt\b^{bn}+m\b^{am}n\b^{bn})
(F(\hHa))_{am}{}^{cp}\het_{cp,bn}
+{k\over 2}\pt\b^{am}(G(\hHa))_{am}{}^{cp}\het_{cp,bn}in\b^{bn}
\nonumber \\ \label{La}\\
&=&{k\over 8\p}\int d\s \left[ (\pt\b^a\pt\b^b-\ps\b^a\ps\b^b)
(F(\Ha))_a{}^c\et_{cb}
+2\pt\b^{a}(G(\Ha))_{a}{}^c\et_{cb}\ps\b^{b} \right]
\nonumber \\ \label{Lb}\\
&=&{k\over8\p}\int d\s \left[
\et_{ab}e_i{}^a e_j{}^b (\pt x^i\pt x^j - \ps x^i \ps x^j)
 + 2B_{ij}\pt x^i \ps x^j \right] \label{Lc}
=L_{WZW}.
\end{eqnarray}
\end{subequations}
To obtain (\ref{La}) from (\ref{Bform3}),
we used the explicit $\b$-basis form of the
mechanical action given in Appendix B\@.  This Appendix also gives the
functions $F(H)$ and $G(H)$.  The form in (\ref{Lb}) then follows from
the mode relations (\ref{relations}b,c,d).  Finally, one notices that
(\ref{Lb}) is the $\b$-basis form (see Appendix A) of the
conventional WZW
Lagrangian on $G$.
In this computation,
the $(\hO^\ys)^2$ term becomes the $(\ps x)^2$ term
(as can also be seen from the relation (\ref{ommod}))
and the $\hB$ term becomes the $B$ term.

\subsection{WZW in terms of $\hg(x(\t))\in\hat G$}
\label{mechg}

We can also use the translation dictionary to obtain the $\hg$ form
of the mechanical action,
\begin{subequations}
\begin{equation}
S_M=-{k\over4\chi}\int \!d\t \hTr(\hg^{-1}\pt\hg\hg^{-1}\pt\hg
 -\hg^{-1}\hg'\hg^{-1}\hg')
+{k\over2\chi}\int \!d\t \!\int_0^1 \!d\rho \vare^{AB}\hTr(\hg^{-1}\hg'\hg^{-1}
 \pa_{A}\hg\hg^{-1}\pa_{B}\hg)
\label{nana}
\end{equation}
\begin{equation}
(\hg')_{Im}{}^{Jn}\equiv i(n-m)\hg_{Im}{}^{Jn} , \qquad
A=(\t,\rho), \qquad \vare^{01}=+1
\label{g'0}
\end{equation}
\label{hgform}
\end{subequations}
where $\hg(\hT,x(\t))\in\hat G$ is the reduced affine group element
in affine matrix representation $\hT$.
This form of the mechanical action
follows directly from the conventional WZW
action
\begin{eqnarray}
S_{WZW}&=&-{k\over2\p\chi}\int d\t d\s \Tr(g^{-1}\pa g g^{-1}\bpa g)
-{k\over12\p\chi}\int \Tr(g^{-1}dg)^3
\label{convwzw}\nonumber \\
&=& S_M
\label{Seqs}
\end{eqnarray}
using\footnote{For the WZW terms, the mode identity is
$\b^a(\t,\s,\rho)=\sum_m e^{im\s}\b^{am}(\t,\rho)$.
Regularity of the conventional WZW term requires $\ps\b^a(\rho=0)=0$
along the axis of the cylinder,
which implies that $m\b^{am}(\rho=0)=0$ on $\hat G$.  It follows that
$W^A(\rho=0)=0$, where $W^A$ is defined in (\ref{Wdef}).
This is why there is no boundary term at $\rho=0$ in (\ref{answer}).}
 the $g\leftrightarrow\hg$ mode
relation (\ref{gmod}) and the trace relation (\ref{trmod}).

The elegant action (\ref{hgform}) is another central result of this paper.
This form of the mechanical action shows a two-dimensional WZW term on
$\hat G$, which is equal, under the translation dictionary,
to the conventional three-dimensional WZW term on $G$.

One can say more about the structure of the two-dimensional WZW term.
Comparing (\ref{Bform3}) and (\ref{nana}), we find
the identity
\begin{equation}
\int d\t \!\int_0^1 d\rho \vare^{AB} \hTr(\hg^{-1}\hg'\hg^{-1}
 \pa_{A}\hg\hg^{-1}\pa_{B}\hg)
=\chi\int d\t \pt x^{i\m}\hB_{i\m,j\n}e_{am}{}^{j\n}\et^{ab}\hO_{b,-m}{}^\ys
\end{equation}
between the two-dimensional and the one-dimensional forms of the WZW
term on $\hat G$.
This identity also follows from the divergence relation
\begin{subequations}
\begin{equation}
\vare^{AB}\hTr(\hg^{-1}\hg'\hg^{-1} \pa_{A}\hg\hg^{-1}\pa_{B}\hg)
= \pa_A W^A
\end{equation}
\begin{equation}
W^A\equiv -\chi\vare^{AB} \pa_B x^{i\m} \hB_{i\m,j\n}
e_{am}{}^{j\n}\et^{ab}\hO_{b,-m}{}^\ys
\label{Wdef}
\end{equation}
\begin{equation}
\int \!d\t \!\int_0^1 \!d\rho \vare^{AB} \hTr(\hg^{-1}\hg'\hg^{-1}
 \pa_{A}\hg\hg^{-1}\pa_{B}\hg)
=\!\int \!d\t W^\rho(\t,\rho=1)
=\chi\int \!d\t \pt x^{i\m}\hB_{i\m,j\n}e_{am}{}^{j\n}\et^{ab}\hO_{b,-m}{}^\ys
\label{answer}
\end{equation}
\end{subequations}
whose structure parallels that of
the WZW term on $\hat G$
in (\ref{diverg})
and the conventional WZW term on $G$.
For a direct proof of the divergence relation, follow the steps,
\begin{subequations}
\begin{eqnarray}
\vare^{AB}\hTr(\hg^{-1}\hg'\hg^{-1}\pa_{A} \hg \hg^{-1}\pa_{B} \hg) &=&
{1\over2\p}\int d\s \Tr(g^{-1}\pa_\s g [g^{-1}\pa_{\t} g ,g^{-1}\pa_{\rho} g])
\label{stepa}
\\ &=&
{\chi\over2\p}\int d\s f_{ab}{}^c
\pa_\t x^i \pa_\rho x^j \pa_\s x^k e_i{}^a e_j{}^b e_k{}^d \et_{dc}
\label{stepb} \\ &=&
\chi f_{am,bn}{}^{cp}
\pa_\t x^{i\m} \pa_\rho x^{j\n}e_{i\m}{}^{am}e_{j\n}{}^{bn}
\hO_{cp}{}^\ys
\label{stepc}\\ &=&
-\chi\vare^{AB}\pa_B x^{i\m}\pa_A(\hB_{i\m,j\n}e_{am}{}^{j\n}\et^{ab}
\hO_{b,-m}{}^\ys)
\label{stepd}
\\ &=&
\pa_A W^A .
\end{eqnarray}
\end{subequations}
To obtain (\ref{stepa}) one uses the $g\leftrightarrow\hg$ mode relation
(\ref{gmod}) and the trace relation (\ref{trmod}), as above.  The form
in (\ref{stepb}) follows by evaluation of the trace.  To obtain
(\ref{stepc}), one returns to $\hat G$ by the mode relation
(\ref{FFF}) (or the vielbein relation (\ref{vielmod})) and the
explicit $\b$-basis form of the adjoint action
$\hO$ in Appendix B\@.
The $\hB$ form in (\ref{stepd}) is then obtained from (\ref{Bident}).

We have also worked out the variation of the mechanical action (\ref{hgform}),
using
\begin{equation}
\d(\hg')_{Im}{}^{Jn}=i(n-m)(\d\hg)_{Im}{}^{Jn}.
\end{equation}
For this computation, it is useful to define the prime operation
on any matrix-valued function,
\begin{equation}
(\F')_{Im}{}^{Jn}\equiv i(n-m)\F_{Im}{}^{Jn}
\end{equation}
which includes (\ref{g'0}) as a special case.
The prime operation satisfies a Leibnitz rule and an identity which is
analogous to integration by parts,
\begin{subequations}
\begin{equation}
(\F_1\F_2)'=\F_1'\F_2+\F_1\F_2'
\end{equation}
\begin{equation}
\hTr(\F_1'\F_2)=-\hTr(\F_1\F_2').
\end{equation}
\end{subequations}
Then we find that the variation
of the WZW term is a total derivative,
\begin{equation}
\d\left(\vare^{AB}\hTr(\hg^{-1}\hg'\hg^{-1}\pa_A\hg\hg^{-1}\pa_B\hg)\right) =
\pa_A\left(\vare^{AB}\hTr(\hg^{-1}\d\hg[\hg^{-1}\pa_B\hg,\hg^{-1}\hg'])\right)
\end{equation}
and the resulting equation of motion of the mechanical system is
\begin{equation}
\pt(\hg^{-1}\pt\hg+\hg^{-1}\hg') -(\hg^{-1}\pt\hg+\hg^{-1}\hg')' = 0.
\label{hEOM}
\end{equation}
Under the translation dictionary, the prime operation becomes the
derivative with respect to $\s$, and the equation of motion (\ref{hEOM})
becomes the usual equation of motion
$\bpa(g^{-1}\pa g)=0$ of the conventional WZW formulation on $G$.

\subsection{Formal Haar measure on $\hat G$ and formal quantum equivalence}
\label{Haar}

In this section, we give the relation between the formal Haar measure on the
affine group and the Haar measure on the Lie group.  We use this
relation to establish the formal quantum equivalence of the mechanical
formulation on $\hat G$ with the conventional WZW formulation on $G$.
The statements of this section hold only up to irrelevant constants.

The formal Haar measure
$d\hg$ on $\hat G$ is equal, under the translation dictionary, to the
spatial product of Haar measures $dg$ on $G$,
\begin{subequations}
\begin{equation}
d\hg(x)=\prod_\s dg(x(\s))
\label{haarequiv}
\end{equation}
\begin{equation}
d\hg(x)\equiv\left(\prod_{i,\m} dx^{i\m}\right)
\sqrt{\det \hat G(x)}, \qquad
dg(x(\s))\equiv\left(\prod_i dx^i(\s)\right)
\sqrt{\det G(x(\s))}
\end{equation}
\begin{equation}
\hat G_{i\m,j\n}\equiv e_{i\m}{}^{am}\het_{am,bn}e_{j\n}{}^{bn}, \qquad
G_{ij}\equiv e_i{}^a \et_{ab} e_j{}^b
\label{Gdef}
\end{equation}
\end{subequations}
where $\hat G_{i\m,j\n}$ and $G_{ij}$ are the target-space metrics on
$\hat G$ and $G$.

Our proof of (\ref{haarequiv}) goes through tangent-space variables
as usual,
\begin{subequations}
\begin{eqnarray}
d\hg(x)&=&\left(\prod_{a,m}d\b^{am}(x)\right)
  e^{{1\over2}(\hTr\ln \hat\G)
    (\sum_m)}
\\
&=&\left(\prod_{a,\s}d\b^a(x(\s))\right)
  e^{{1\over2}\d(0)\int d\s\Tr\ln\G}
\label{haarb}
\\
&=&\prod_\s dg (x(\s))
\end{eqnarray}
\begin{equation}
\hat\G_{am}{}^{bn}\equiv \pa_{am}x^{i\m}\hat G_{i\m,j\n}\pa_{cp}x^{j\n}
\het^{cp,bn}, \qquad
\G_a{}^b\equiv \pa_a x^i G_{ij}\pa_c x^j\et^{cb}
\end{equation}
\end{subequations}
where $\d(0)=(1/2\p)\sum_m$ is the periodic delta function $\d(\s)$ at
$\s=0$.

Taken with the action equality $S_M=S_{WZW}$ in (\ref{Seqs}),
this result shows the formal quantum equivalence of the mechanical
formulation on $\hat G$ with the conventional formulation
of WZW on $G$,
\begin{subequations}
\begin{equation}
\int (\cD_M \hg)\, e^{iS_M} =
\int (\cD g)\, e^{iS_{WZW}}
\label{MW}
\end{equation}
\vskip -13pt
\begin{equation}
\cD_M \hg \equiv \prod_\t d\hg(x(\t)), \qquad
\cD g \equiv \prod_{\t,\s} dg(x(\t,\s))
\label{functmeas}
\end{equation}
\begin{equation}
\cD_M\hg=\cD g
\end{equation}
\label{morestuff}
\end{subequations}
where $\cD g$ in (\ref{functmeas}) is the formal functional measure of
the conventional formulation.
It follows from (\ref{morestuff}) and (\ref{Fourb}) that
conventional WZW averages on $G$
\begin{equation}
\langle \F[\b^a(x(\t,\s))]\rangle\raisebox{-2pt}{${}_G$} =
\langle \F[\sum_m e^{im\s}\b^{am}(x(\t))]\rangle_{\hat G_M}
\label{ggtrans2}
\end{equation}
can be formally computed for any $\F$ as shown, using the mechanical
formulation ($\hat G_M$) on $\hat G$.

As seen above, the formal measures $d\hg(x)=\prod_\s dg(x(\s))$ and
$\cD_M\hg=\cD g$ have closely related formal divergences.
It is an important open problem to find suitably regularized forms of
these measures.

\section{Field theory on $\hat G$ and formal quantum equivalence}
\label{Funmeas}

In this section we use the formal Haar measure on $\hat G$ to discuss the
formal quantum equivalence
\begin{equation}
\begin{picture}(200,90)(0,80)
\put(5,150){$\hat g(x(\tau))$}
\put(0,80){$g(x(\tau,\sigma))$}
\put(45,153){\vector(1,0){80}}
\put(50,153){\vector(-1,0){5}}
\put(130,150){$\hat g(x(\tau,\sigma))$}
\put(50,98){\vector(3,2){75}}
\put(50,98){\vector(-3,-2){12}}
\put(125,80){$g(x(\tau,\sigma,\ts))$}
\put(60,160){constraint}
\multiput(65,80)(20,0){3}{$\cdot$}
\multiput(150,100)(0,15){3}{$\cdot$}
\multiput(25,100)(0,15){3}{$\cdot$}
\end{picture}
\end{equation}
of the field theory on $\hat G$ with the mechanical system on $\hat G$
and the conventional formulation of WZW on $G$.
These equivalences were discussed at the classical level in Section
\ref{sectt}.
Our conclusion is that we may take  the spacetime
product of Haar measures on $\hat G$ as the
functional measure for the field theory on $\hat G$,
and this measure dictates
a more elegant form for the constraint term of this formulation.
Again, the measure relations of this section hold only up to
irrelevant constants.

For the discussion here, it is convenient to write the action
(\ref{traction}) of the field theory on $\hat G$ as
\begin{subequations}
\begin{equation}
S_{\hat F}=S_0 + S_\C
\end{equation}
\begin{equation}
S_0=-{k\over2\p\chi}\int d\t d\s
\hat{\Tr}\left(\hg^{-1}\pa\hg
\hg^{-1}\bar\pa\hg
 \right)
 - {k\over 12\p \chi} \int \hTr(\hg^{-1}d\hg)^3
\end{equation}
\begin{equation}
S_\C={k\over\chi}\int d\t d\s
\hTr( \hT_{am}\ln \hg)
 (i\ps+m)\l^{am}
\end{equation}
\label{S0}
\end{subequations}
where $S_\C$ is the constraint term.

Starting from the partition function (\ref{MW}) of the
mechanical formulation, we can use the constraint identities
(\ref{meas1}) and (\ref{meas2}) to derive the partition function of
the field theory on $\hat G$.  The result is
\begin{equation}
\int (\cD_M\hg) \,e^{iS_M} =
\int \left(\prod_{\t,\s,i,\m} dx^{i\m}(\t,\s)\right)
\left(\prod_\t\sqrt{\det \hat G(x(\t,\s_0)}\right)
e^{iS_0}\d[\pa_\s x^{i\m}-\et^{ab}e_{am}{}^{i\m}\hO_{b,-m}{}^\ys]
\label{unaesth}
\end{equation}
where $\hat G$ is defined in (\ref{Gdef})
and we used the fact (see eq.(\ref{FMeq}))
that $S_0=S_M$ on the constrained subspace.
The functional measure in this relation
contains an unaesthetic product of affine Haar measures at a fixed
reference point $\s_0$ in $\s$.

We can obtain a more elegant form of the functional
measure by changing variables to the reduced affine
group element $\hg$.  One begins with the solution of the constraint
\begin{equation}
\b^{am}(x(\t,\s))=e^{im\s}\b^{am}(x(\t))
\end{equation}
given in (\ref{bconst}).  Following steps parallel to those
used in proving eq.(\ref{uhhuh}), we find the $\s$-dependence of
$\hg$
\begin{subequations}
\begin{equation}
\hg(\hT,x(\t,\s))_{Im}{}^{Jn}= e^{i\b^{am}(x(\t,\s))\hT_{am}}
= e^{i(n-m)\s}\hg(\hT,x(\t))_{Im}{}^{Jn}
\end{equation}
\begin{equation}
\ps\hg=\hg'
\end{equation}
\begin{equation}
(\hg')_{Im}{}^{Jn}\equiv
i(n-m)\hg_{Im}{}^{Jn}
\label{g'}
\end{equation}
\end{subequations}
on the constrained subspace.

One may then prove the following relation
\begin{equation}
\hg^{-1}(\ps\hg-\hg')=i(\ps x^{i\m}-\et^{ab}e_{am}{}^{i\m}\hO_{b,-m}{}^\ys)
 e_{i\m}{}^{cp}\hT_{cp}
\label{provethis}
\end{equation}
which holds on or off the constrained subspace, and which gives the $\hg$
form of the constraint.
This relation can be proven order by order in $\b^{am}$,
or by the following simple argument.
Using only chain rule and the vielbein relation (\ref{Tviel}), one
finds that $\hg(x(\s))$ satisfies
\begin{equation}
\hg^{-1}\ps\hg=i\ps x^{i\m} e_{i\m}{}^{am}\hT_{am}
\end{equation}
on or off the constrained subspace.  Because
both sides of (\ref{provethis}) vanish on the constrained subspace,
it then follows that
\begin{equation}
\hg^{-1}\hg'=i\et^{ab}\hO_{a,-m}{}^{\ys}\hT_{bm}
\label{heylook}
\end{equation}
on the constrained subspace.  But neither side of (\ref{heylook}) has any
sigma derivatives, so this relation, and hence (\ref{provethis}), must be
true on or off the constrained subspace.

Using the relation (\ref{provethis}) in the
partition function (\ref{unaesth}), we find that
\begin{subequations}
\begin{equation}
\int(\cD_M\hg)\, e^{iS_M}=
\int(\cD_F\hg)\, e^{iS_0}\d[\hg^{-1}(\ps\hg-\hg')]=
\int(\cD\l\cD_F\hg )\, e^{iS_F}
\label{MFF}
\end{equation}
\begin{equation}
\cD_F \hg\equiv \prod_{\t,\s} d\hg(x(\t,\s))
\end{equation}
\end{subequations}
where the $\hg$ form of the constraint appears in the functional
delta function, and
$\cD_F\hg$ is the spacetime product of Haar measures on $\hat G$.
The improved action $S_F$ of the field theory on $\hat G$ is
\begin{eqnarray}
S_F&=&-{k\over2\p\chi}\int d\t d\s
\hat{\Tr}\left(\hg^{-1}\pa\hg
\hg^{-1}\bar\pa\hg
 \right)
 - {k\over 12\p \chi} \int \hTr(\hg^{-1}d\hg)^3
\nonumber \\
& &
 +{k\over\chi}\int d\t d\s
\hTr( \l\hg^{-1}(\ps \hg-\hg'))
\label{impraction}
\end{eqnarray}
where $\hg(\hT,x(\t,\s))\in\hat G$ is the reduced affine group element
and $\hg'$ is defined in eq.(\ref{g'}).
Because of the formal quantum equivalence (\ref{MFF}),
this form of the field theory on $\hat G$ is  another central
result of the paper.
Of course, the improved action $S_F$ and the action
$S_{\hat F}=S_0+S_\C$ in (\ref{S0}) are classically
equivalent, differing only in the form
of the constraint.

\vskip .2cm \noindent
\underline{Summary of the quantum equivalences}

Collecting the results (\ref{MW}) and (\ref{MFF}), we have the formal
quantum equivalences,
\begin{equation}
\int(\cD g)\, e^{iS_{WZW}}=
\int(\cD_M\hg)\, e^{iS_M}=
\int(\cD\l\cD_F\hg)\, e^{iS_F}
\end{equation}
among all three formulations of WZW theory.  Similarly, the results
(\ref{hghgtrans}) and (\ref{ggtrans2}) may be combined to obtain the
relations
\begin{equation}
\langle \F[\b^a(x^i(\t,\s))]\rangle\raisebox{-2pt}{${}_G$} =
\langle \F[\sum_m e^{im\s}\b^{am}(x^{i\m}(\t))]\rangle_{\hat G_M} =
\langle \F[\sum_m \b^{am}(x^{i\m}(\t,\s))]\rangle_{\hat G_F}
\end{equation}
among the formal averages of all three formulations.

\section{WZW as a field theory in three dimensions}
\label{wzw3}

We finally turn to WZW as a three-dimensional
field theory on $G$, a formulation
which we will derive from the two-dimensional
field theory on $\hat G$
\begin{equation}
\begin{picture}(200,90)(0,80)
\put(5,150){$\hat g(x(\tau))$}
\put(0,80){$g(x(\tau,\sigma))$}
\put(130,150){$\hat g(x(\tau,\sigma))$}
\put(150,140){\vector(0,-1){50}}
\put(150,135){\vector(0,1){5}}
\put(125,80){$g(x(\tau,\sigma,\ts))$}
\put(113,115){modes}
\multiput(65,80)(20,0){3}{$\cdot$}
\multiput(65,150)(20,0){3}{$\cdot$}
\multiput(25,100)(0,15){3}{$\cdot$}
\end{picture}
\end{equation}
by another application of the translation dictionary of Section
\ref{modestolocal}.  In this case, the partial transmutation of the
base and target space is
\begin{subequations}
\begin{equation}
x^{i\m}(\t,\s)\leftrightarrow x^i(\t,\s,\ts)
\end{equation}
\begin{equation}
\hg(x(\t,\s)): \;\; \R\times S^1\mapsto \hat G, \qquad\;\;
g(x(\t,\s,\ts)): \;\; \R\times T^2 \mapsto G
\end{equation}
\end{subequations}
where $0\le\ts<2\p$ is an additional periodic coordinate.

We begin with the form of the two-dimensional field-theory on $\hat G$,
\begin{eqnarray}
S&=&-{k\over2\p\chi}\int d\t d\s
\hat{\Tr}\left(\hg^{-1}\pa\hg
\hg^{-1}\bar\pa\hg
 \right)
 - {k\over 12\p \chi} \int \hTr(\hg^{-1}d\hg)^3
\nonumber \\
& &
 +{k\over\chi}\int d\t d\s
\hTr( \l\hg^{-1}(\ps\hg-\hg'))
\label{traction2}
\end{eqnarray}
obtained in Section \ref{Funmeas}.
Next, we define fields on $G$ which are local in $\ts$,
\begin{subequations}
\begin{equation}
\b^a(x(\t,\s,\ts))\equiv\sum_m e^{im\ts} \b^{am}(x(\t,\s)), \qquad
(\l(\t,\s,\ts))_I{}^J\equiv\sum_m e^{i(n-m)\ts} (\l(\t,\s))_{Im}{}^{Jn}
\label{sigprime}
\end{equation}
\begin{equation}
g(T,x(\t,\s,\ts))\equiv e^{i\b^a(x(\t,\s,\ts))T_a} \in G.
\end{equation}
\end{subequations}
Then the action (\ref{traction2}) can be reexpressed as a three-dimensional
field theory on $G$,
\begin{subequations}
\begin{eqnarray}
S_3&=&-{k\over4\p^2\chi}\int d\t d\s d\ts \Tr \left(g^{-1}\pa g
 g^{-1}\bar\pa g\right) - {k\over 24\p^2 \chi} \int \Tr(g^{-1} dg)^3
\!\wedge\! d\ts
\nonumber \\
& &
 +{k\over2\p\chi}\int d\t d\s d\ts
 \Tr( \l g^{-1}(\ps-\pa_{\ts})g )
\label{3d}
\end{eqnarray}
\begin{equation}
\Tr(g^{-1}dg)^3 \!\wedge\! d\ts =
d\t d\s d\rho d\ts
\vare^{ABC}\Tr(g^{-1}\pa_A gg^{-1}\pa_B g g^{-1}\pa_C g)
\end{equation}
\begin{equation}
A=(\t,\s,\rho), \qquad \vare^{012}=+1
\end{equation}
\label{3d2}
\end{subequations}
with a four-dimensional WZW term.
The first two terms of (\ref{3d}) follow from (\ref{traction2})
in a single step using the trace identity
(\ref{trmod}), and the
form of the constraint term follows from (\ref{trmod}) and (\ref{sigprime}).

The three-dimensional action (\ref{3d2}) completes the quartet of
formulations of WZW theory
announced in the introduction and shown in eq.(\ref{picture}).

It is clear from the derivation above that the three-dimensional
 form of WZW theory is
equivalent to the other three formulations.  Instead
of developing translation dictionaries between this formulation and
the formulations on $\hat G$, we confine ourselves here to showing
a direct equivalence with the conventional WZW formulation on $G$,
\begin{equation}
\begin{picture}(200,90)(0,80)
\put(5,150){$\hat g(x(\tau))$}
\put(0,80){$g(x(\tau,\sigma))$}
\put(130,150){$\hat g(x(\tau,\sigma))$}
\put(125,80){$g(x(\tau,\sigma,\ts))$}
\put(52,83){\vector(1,0){69}}
\put(62,83){\vector(-1,0){10}}
\put(60,90){constraint}
\multiput(65,150)(20,0){3}{$\cdot$}
\multiput(25,100)(0,15){3}{$\cdot$}
\multiput(150,100)(0,15){3}{$\cdot$}
\end{picture}
\label{ashown}
\end{equation}
in parallel with the demonstration at the end of Section \ref{sectt}.

To see this equivalence, we start with the three-dimensional formulation
and solve its constraint,
\begin{equation}
(\ps-\pa_{\ts})g(x(\t,\s,\ts))=0
\end{equation}
which is
the equation of motion of the multiplier $\l$ in (\ref{3d}).
This constraint can be simplified to
\begin{equation}
(\ps-\pa_{\ts})x^i(\t,\s,\ts)=0
\label{ssconst}
\end{equation}
and one also finds that
\begin{equation}
\b(x(\t,\s,\ts))=e^{im(\s+\ts)}\b^{am}(x(\t))
\label{newbeta}
\end{equation}
by using and eqs.(\ref{sigprime}) and (\ref{ssconst}).

Then  it is convenient to define new variables on $T^2$,
\begin{equation}
\s^+\equiv\s+\ts, \qquad \s^-\equiv\s-\ts, \qquad
0\le\s^-<2\p
\end{equation}
so that $x^i=x^i(\t,\s^+)$.
It follows from periodicity on $T^2$
that, as in (\ref{newbeta}), all
the quantities of the theory on the constrained subspace
are periodic functions $f_{2\p}(x(\t,\s^+))$ of $\s^+$
with period $2\p$, and moreover,
\begin{equation}
\int_0^{2\p} d\s \int_0^{2\p} d\ts f_{2\p}(x(\t,\s^+)) =
\int_0^{2\p} d\s^- \int_0^{2\p} d\s^+ f_{2\p}(x(\t,\s^+)).
\end{equation}
Therefore, on the constrained subspace, the three-dimensional
action (\ref{3d}) has the form
\begin{equation}
S=-{k\over2\p\chi}\int d\t d\s^+ \Tr \left(g^{-1}\pa g
 g^{-1}\bar\pa g\right) - {k\over 12\p \chi} \int \Tr(g^{-1} dg)^3
\label{2d}
\end{equation}
after doing the integration over $\s^-$.  The three-form in
(\ref{2d}) now contains a factor
$d\t d\s^+ d\rho$ so, with the identification
$\s_{WZW}\equiv\s^+$,   this is the conventional WZW action on $G$.

\addcontentsline{toc}{section}{Acknowledgements}
\section*{Acknowledgements}

We thank
N. Arkani-Hamed, J. deBoer, P. Bouwknegt,
V. Jones, H. Ooguri, R. Plesser, N. Sochen,
and W. Taylor for helpful discussions.

This work was
supported in part by the Director, Office of
Energy Research, Office of High Energy and Nuclear Physics, Division of
High Energy Physics of the U.S. Department of Energy under Contract
DE-AC03-76SF00098 and in part by the National Science Foundation under
grant PHY90-21139.

\appendix
\addcontentsline{toc}{section}{Appendix A:$\;$  Identities on $G$}
\section*{Appendix A:$\;$  Identities on $G$}
\def\theequation{A.\arabic{equation}}
\setcounter{equation}{0}
\label{lapp}

We list below some useful identities on the Lie group $G$, including
in particular
 the explicit $\b$-basis forms of various quantities
which are central to the proofs of Sections \ref{modestolocal},
\ref{lflc} and \ref{WZWequiv}.
Here, $J_a, a=1\ldots\dim g$
are the generators of Lie $g$, and $x^i, i=1\ldots\dim g$
are Einstein coordinates on the group manifold.

\noindent A. Group element and adjoint action.
\begin{subequations}
\begin{equation}
g(J,x)=e^{i\b^a(x)J_a}
\end{equation}
\begin{equation}
g J_a g^{-1} = \O_a{}^b J_b, \qquad
\O_a{}^c \et_{cd} \O_b{}^{d} = \et_{ab}
\end{equation}
\begin{equation}
\O_a{}^b = \left( e^{-i H^{adj}}\right)_a{}^b, \qquad
H^{adj}=\b^a T_a^{adj}.
\end{equation}
\end{subequations}
The quantities $\et_{ab}$ and $T_a^{adj}$ are
the Killing metric and adjoint representation of Lie $g$.

\noindent B. Vielbeins and inverse vielbeins.
\begin{subequations}
\begin{equation}
e_i = -i g^{-1} \pa_i g = e_i{}^a J_a, \qquad
\bar e_i = -i g \pa_i g^{-1} = \bar e_i{}^a J_a
\end{equation}
\begin{equation}
\pa_i e_{j}{}^a - \pa_j e_i{}^a = e_i{}^b e_j{}^c f_{bc}{}^a, \qquad
e_a{}^i \pa_i e_b{}^j - e_b{}^i \pa_i e_a{}^j =
f_{ba}{}^c e_c{}^j
\label{CMICM}
\end{equation}
\begin{equation}
\bar e_i{}^a = -e_i{}^b \O_b{}^a, \qquad
\bar e_a{}^i = -(\O^{-1})_a{}^b e_b{}^i
\end{equation}
\begin{equation}
e_i{}^a = \pa_i\b^{b} (M(H^{adj}))_b{}^a, \qquad
M(H)={e^{iH}-1 \over iH}.
\end{equation}
\end{subequations}
The Cartan-Maurer and inverse Cartan-Maurer relations in
(\ref{CMICM}) hold also for $e\rightarrow\bar e$.

\noindent C. Antisymmetric tensor field.
\begin{subequations}
\begin{equation}
B_{ij}=\pa_i\b^b \pa_j \b^a \et_{bc} (N(H^{adj}))_a{}^c, \qquad
N(H)={(e^{iH}-e^{-iH})-2iH \over (iH)^2}
\end{equation}
\begin{equation}
B_{ij}={i\over\chi}\int_0^1 dt\, \Tr \left(
H\pa_{[i}e^{-itH}\pa_{j]}e^{itH} \right), \qquad
H=\b^aT_a
\end{equation}
\begin{equation}
\pa_i B_{jk}+\pa_j B_{ki}+\pa_k B_{ij} =
{i \over \chi} \Tr(e_i[e_j,e_k]), \qquad
\Tr(T_a T_b) = \chi \eta_{ab}.
\end{equation}
\end{subequations}
$T_a$ is any matrix irrep of Lie $g$.

\noindent D. Local currents on $G$.
\begin{subequations}
\begin{equation}
E_a(\s)=2\p e_a{}^ip_i(B)+{k\over2}\et_{ab}e_i{}^b\ps x^i
= 2\p (C(H^{adj}))_a{}^b p_b -k\pa_\s \b^c (D(H^{adj}))_c{}^b \eta_{ba}
\end{equation}
\begin{equation}
\bar E_a(\s)=2\p \be_a{}^ip_i(B)-{k\over2}\et_{ab}\be_i{}^b\ps x^i
= 2\p (\bar C(H^{adj}))_a{}^b p_b +k\pa_\s \b^c
(\bar D(H^{adj}))_c{}^b \eta_{ba}
\end{equation}
\begin{equation}
p_i(B)= p_i-{k\over4\p}B_{ij}\ps x^j, \qquad
p_i=\pa_i \b^a p_a
\end{equation}
\begin{equation}
C(H)= {i H \over e^{iH}-1}, \qquad
D(H)= {e^{-iH}-1+iH \over iH(e^{-iH}-1)}
\end{equation}
\begin{equation}
\bar C(H)= {i H \over e^{-iH}-1}, \qquad
\bar D(H)= {e^{iH}-1-iH \over iH(e^{iH}-1)}.
\end{equation}
\end{subequations}
The canonical momenta $p_i(\s)$ and $p_a(\s)$, which can
be classical or quantum, are defined in Section
\ref{modestolocal}.

\noindent E. Conventional WZW Lagrangian on $G$
\begin{subequations}
\begin{eqnarray}
L_{WZW}
&=&{k\over8\p}\int d\s \left[
\et_{ab}e_i{}^a e_j{}^b (\pt x^i\pt x^j - \ps x^i \ps x^j)
 + 2B_{ij}\pt x^i \ps x^j \right] \\
&=&{k\over 8\p}\int d\s \left ((\pt\b^a\pt\b^b-\ps\b^a\ps\b^b)
(F(\Ha))_a{}^c\et_{cb}
+2\pt\b^{a}(G(\Ha))_{a}{}^c\et_{cb}\ps\b^{b} \right)
\nonumber \\
\end{eqnarray}
\begin{equation}
F(H)={2-e^{iH}-e^{-iH}\over H^2}, \qquad
G(H)=N(H)={(e^{iH}-e^{-iH})-2iH\over (iH)^2}.
\end{equation}
\end{subequations}
The $g$ form of the conventional WZW action is given in (\ref{wzwaction}).

\addcontentsline{toc}{section}{Appendix B:$\;$ Identities on $\hat G$}
\section*{Appendix B:$\;$ Identities on $\hat G$}
\def\theequation{B.\arabic{equation}}
\setcounter{equation}{0}
\label{aapp}

We list some useful identities on the affine Lie group
$\hat G$ (analogous to those on $G$
in Appendix A)
including in particular the explicit $\b$-basis
forms of various quantities which are central to the proofs of
Section \ref{modestolocal}, \ref{lflc} and \ref{WZWequiv}.
Here $J_a(m), a=1\ldots\dim g, m\in \Z$ are the current modes
and $x^{i\m}, i=1\ldots\dim g, \m\in\Z$ are the coordinates on the
reduced affine group manifold.
Einstein and tangent-space indices are $\L,\Ga=(i\m,y)$ and
$L,M=(am,\ys)$ respectively, and $\J_L=(J_a(m),k)$ are the generators of
the affine group.

\noindent A.  Reduced group element and adjoint action.
\begin{subequations}
\begin{equation}
\hg(J,x)=e^{i\b^{am}(x)J_a(m)}
\end{equation}
\begin{equation}
\hg\J_L\hg^{-1}=\hO_L{}^M\J_M, \qquad
\hO_{am}{}^{cp}\het_{cp,dq}\hO_{bn}{}^{dq} = \het_{am,bn}
\end{equation}
\begin{equation}
\hO_L{}^M = \left( e^{-i\hHa}\right)_L{}^M, \qquad
\hHa=\b^{am}\hT_{am}^{adj}
\end{equation}
\begin{equation}
\hO_\ys{}^L=\d_\ys{}^L, \qquad
\hO_{am}{}^\ys=\left({e^{-i\hHa}-1\over \hHa }
\right)_{am}^{~~~\raisebox{-5pt}{\scriptsize$bn$}}
(\hHa)_{bn}{}^\ys.
\end{equation}
\end{subequations}
The quantities
$\het_{am,bn}$ and $\hT^{adj}$ are the rescaled Killing
metric (see Section \ref{traces}) and the
adjoint representation of the affine algebra.

\noindent B.  Vielbeins and inverse vielbeins.
\begin{subequations}
\begin{equation}
e_{i\m}=-i\hg^{-1}\pa_{i\m}\hg=e_{i\m}{}^L\J_L  , \qquad
\be_{i\m}=-i\hg\pa_{i\m}\hg^{-1}=\be_{i\m}{}^L\J_L
\end{equation}
\begin{equation}
\pa_{i\m}e_{j\n}{}^{L}-\pa_{j\n}e_{i\m}{}^{L}
= e_{i\m}{}^{M}e_{j\n}{}^{N}f_{MN}{}^{L}, \qquad
e_L{}^{i\m}\pa_{i\m}e_M{}^{j\n}-e_M{}^{i\m}\pa_{i\m}e_L{}^{i\m} =
f_{ML}{}^N e_N{}^{j\n}
\label{affCM}
\end{equation}
\begin{equation}
\be_{i\m}{}^L=-e_{i\m}{}^M\hO_M{}^L, \qquad
\be_L{}^{i\m}=-(\hO^{-1})_L{}^M e_M{}^{i\m}
\end{equation}
\begin{equation}
e_{i\m}{}^L(x)=\pa_{i\m}\b^{am}(x)(M(\hHa))_{am}{}^L, \qquad
M(H)={e^{iH}-1 \over iH}.
\end{equation}
\end{subequations}
The Cartan-Maurer and inverse Cartan-Maurer relations in (\ref{affCM})
hold also for $e\rightarrow\bar e$.

\noindent C.  Antisymmetric tensor field.
\begin{subequations}
\begin{equation}
\hB_{i\m,j\n}=\pa_{i\m}\b^{bn}\pa_{j\n}\b^{am}\et_{bc}
(N(\hHa))_{am}{}^{c,-n},
\qquad
N(H)={e^{iH}-e^{-iH}-2iH \over (iH)^2}
\end{equation}
\begin{equation}
e_{i\m}{}^\ys={1\over2}\left(\hB_{i\m,j\n}e_{a,-m}{}^{j\n}\et^{ab}
             -e_{i\m}{}^{bm}\right)\hO_{bm}{}^\ys
\end{equation}
\begin{equation}
\hB_{i\m,j\n}={i\over\chi}\int_0^1 dt\, \hTr \left(
\hH\pa_{[i\m}e^{-it\hH}\pa_{j\n]}e^{it\hH} \right), \qquad
\hH=\b^{am}\hat T_{am}
\end{equation}
\begin{equation}
\pa_{i\m}\hB_{j\n,k\rho} +
\pa_{j\n}\hB_{k\rho,i\m} + \pa_{k\rho}\hB_{i\m,j\n} =
{i\over\chi}\hTr(e_{i\m}[e_{j\n},e_{k\rho}])
\end{equation}
\begin{equation}
\pa_{i\m}(\hB_{j\n,k\r}e_{a,-m}{}^{k\r}\et^{ab}\hO_{bm}{}^\ys) -
\left( i\m \lra j\n \right) =
e_{i\m}{}^{bn}e_{j\n}{}^{am}f_{am,bn}{}^{cp}\hO_{cp}{}^\ys.
\end{equation}
\end{subequations}
$\hT$ and $\hTr$ are the matrix representations of
the affine algebra and the reduced traces discussed in Section \ref{traces}.

\noindent D. Affine Lie derivatives.
\begin{subequations}
\begin{equation}
E_a(m)=e_{am}{}^{i\m} p_{i\m}(\hB) + {1\over2}k\hO_{am}{}^\ys
= (C(\hHa))_{am}{}^{bn} p_{bn}
-i k n\b^{bn} (D(\hHa))_{bn}{}^{cp} \het_{cp,am}
\end{equation}
\begin{equation}
\bE_a(m)=\be_{am}{}^{i\m} p_{i\m}(\hB) - {1\over2}k\hO_{am}{}^\ys
= (\bar C(\hHa))_{am}{}^{bn} p_{bn}
+i k n\b^{bn} (\bar D(\hHa))_{bn}{}^{cp} \het_{cp,am}
\end{equation}
\begin{equation}
p_{i\m}(\hB)\equiv p_{i\m} -
{1\over2}k\hB_{i\m,j\n}e_{a,-m}{}^{j\n}\et^{ab}\hO_{bm}{}^\ys, \qquad
p_{i\m}=\pa_{i\m}\b^{am}p_{am}
\end{equation}
\begin{equation}
C(H)= {i H \over e^{iH}-1}, \qquad
D(H)= {e^{-iH}-1+iH \over iH(e^{-iH}-1)}
\end{equation}
\begin{equation}
\bar C(H)= {i H \over e^{-iH}-1}, \qquad
\bar D(H)= {e^{iH}-1-iH \over iH(e^{iH}-1)}.
\end{equation}
\end{subequations}
The canonical momenta $p_{i\m}$ and $p_{am}$, which can be classical
or quantum, are defined in Sections \ref{brackrep} and
\ref{modestolocal}.

\noindent E. Mechanical Lagrangian on $\hat G$.

\begin{subequations}
\begin{eqnarray}
L&=&{k\over4}\et_{ab}e_{i\m}{}^{a,-m}e_{j\n}{}^{bm} \pt x^{i\m}\pt x^{j\n}
  - {k\over4} \left( \hO_{am}{}^\ys+2\hB_{i\m,j\n}
    \pt x^{j\n}e_{am}{}^{i\m}\right)\et^{ab}\hO_{b,-m}{}^\ys \\
&=&{k\over 4}(\pt\b^{am}\pt\b^{bn}+m\b^{am}n\b^{bn})
(F(\hHa))_{am}{}^{cp}\het_{cp,bn}
+{k\over 2}\pt\b^{am}(G(\hHa))_{am}{}^{cp}\het_{cp,bn}in\b^{bn}
\nonumber \\
\end{eqnarray}
\begin{equation}
F(H)={2-e^{iH}-e^{-iH}\over H^2}, \qquad
G(H)=N(H)={(e^{iH}-e^{-iH})-2iH\over (iH)^2}.
\end{equation}
\end{subequations}
The purely group-theoretic forms of the mechanical action on $\hat G$
are given in
eqs.(\ref{myaction}) and (\ref{hgform}).

\end{document}